\documentclass[aps,prl,twocolumn,showpacs,superscriptaddress]{revtex4-1}
\usepackage{amsmath,amssymb}
\usepackage{graphics,graphicx,color}
\usepackage{dcolumn,bm}
\usepackage[normalem]{ulem}
\usepackage{subcaption}
\usepackage{tikz}
\usepackage[colorlinks=true,citecolor=blue,urlcolor=blue]{hyperref}
\usepackage{cases}
\usepackage{soul}
\def\[{\left[}
\def\]{\right]}
\def\({\left(}
\def\){\right)}
\def\be{\begin{equation}}
\def\ee{\end{equation}}
\def\bea{\begin{eqnarray}}
\def\eea{\end{eqnarray}}

\newcommand{\eee}{\mathrm{e}}

\newcommand{\C}{\mathcal{C}}
\newcommand{\Cp}{\mathcal{C'}}
\newcommand{\Cpp}{\tilde{\C}}
\newcommand{\WW}{\mathbb{W}}
\newcommand{\aux}{^{\rm aux}}

\newcommand{\g}{s}
\newcommand{\dt}{\Delta t}
\DeclareSymbolFont{bbold}{U}{bbold}{m}{n}
\DeclareSymbolFontAlphabet{\mathbbold}{bbold}
\newcommand{\mident}{\mathbbold{1}}
\newcommand{\proj}{\langle e|}
\newcommand{\Pf}{P_{\rm f}}

\newcommand{\gaug}
{\affiliation{Institute for Theoretical Physics, Georg-August-Universit\"at G\"ottingen, 37077 G\"ottingen, Germany}}
\newcommand{\kcl}{\affiliation{Department of Mathematics, King's College London, London WC2R 2LS, UK}}

\begin{document}
%\title{Bringing together two paradigms of non-equilibrium:\\Driven dynamics of aging systems}
\title{Bringing together two paradigms of non-equilibrium:\\ Fragile versus robust aging in driven {glassy} systems}

\author{Diego Tapias}
\email[Email: ]{diego.tapias@theorie.physik.uni-goettingen.de}
\gaug

\author{Charles Marteau}
\kcl
\affiliation{Department of Physics and Astronomy, University of British Columbia, Vancouver V6T 1Z1, Canada}

\author{Fabi\'an Aguirre-L\'opez}
\kcl
\affiliation{Chair of Econophysics and Complex Systems, École polytechnique, 
91128 Palaiseau Cedex, France}
\affiliation{LadHyX UMR CNRS 7646, École polytechnique, 91128 Palaiseau Cedex, 
France}

\author{Peter Sollich}%
\email[Email: ]{peter.sollich@uni-goettingen.de}
\gaug
\kcl

\begin{abstract}
There are two {key} paradigms for non-equilibrium dynamics: on the one hand, aging towards an equilibrium state that cannot be reached on reasonable timescales; on the other, external driving that can lead to non-equilibrium steady states. We explore how these two mechanisms interact, by studying the behaviour of trap models, which are paradigmatic descriptions of slow glassy dynamics, when driven by trajectory bias towards high or low activity. To diagnose whether the driven systems continue to age, we establish a framework for mapping the biased dynamics to a Markovian time evolution with time-dependent transition rates. We find that the original aging dynamics reacts in two qualitatively distinct ways to the driving: it can be destroyed by driving of any nonzero strength (\emph{fragile} aging), whereby the dynamics either reaches an active steady state or effectively freezes; or it can persist within a finite range of driving strengths around the undriven case (\emph{robust} aging). This classification into fragile and robust aging could form the basis for distinguishing different universality classes of aging dynamics.
\end{abstract}

\maketitle

%\begin{bibunit}
In non-equilibrium statistical physics there are two {key} paradigms for bringing a system out of equilibrium. The first consists of slowing the dynamics down to such an extent that the system fails to reach equilibrium on practical timescales and instead {\em ages}, with  its properties depending on the time elapsed since preparation. This phenomenon is found in glasses~\cite{KobBar97,LehNag98,KobBarSciTar00,%
SciTar01,MarSchPooPus02,ViaJurLeq03,%
MarBryVan10,WarRot13,%ArcLanBerBir20,
ArcLanBerBir21} after abrupt decreases in temperature or volume, beyond the regime of the supercooled (metastable equilibrium) liquid, but is in fact much more widespread, occurring in many amorphous or disordered materials~\cite{Struik78,%
SibHof89,% 
Bouchaud92,CugKur93,BalCugKurPar95,%
BarMez95,%
BouDea95,%
FraHer95,%
Hodge95,%
Ritort95,%
BouCugKurMez96,%
CugKurLeD96,MonBou96,%
Rieger96,CugDeaKur97,%
VinHamOciBouCug97,%
BouCugKurMez98,%
BerBarKur00,CloBorLei00,%
FieSolCat00,%
RinMaaBou00,%
Kurchan01,%
RamCip01,BenBovGay02,%
CipRam02,%
CasChaCugKen02,BarChe03,BenBovGay03,FieSol03,%
MonRic03,CohHohKhi04,Ritort04b,%
MayLeoBerGarSol06,FieCatSol09,BerBir11,%
SolOliBre17,JosPet18,LubWol18,ParFieSol20,%
ParManSol22}, athermal systems such as crumpled sheets~\cite{LahGotAmiRub17,ShoFriLah23}, active matter~\cite{JanKaiLow17,ManSol20,JanJan21}, cellular protein condensates~\cite{JawFisSahWanFraZha20,TakJawPopJul23}, etc. 

The second {key} paradigm is that of a system brought out of equilibrium by external {\em driving} such as mechanical stress that leads to shear flow, or a magnetic field that creates a current around a ring. 
{Closely related to this is driving in active matter systems, which there arises from e.g.\ self-propulsion forces acting on the constituent particles; indeed recent research 
has revealed some intriguing connections between external driving by shear and driving by activity~\cite{HenFilMar11,NanGov18,LiaXu18,
MoLiaXu20,ManSol20,ManSol21,ManSol21a,
MorRoyAgoStaCorMan21,VilDur21,
KetManSolJacBer23}.}
For steady driving a non-equilibrium steady state is then often reached, which balances energy input and dissipation. Of particular interest within this class of non-equilibrium states have recently been systems driven by trajectory bias, see e.g.~\cite{lebowitz1999gallavotti, evans2004rules, evans2004detailed, lecomte2007thermodynamic, garrahan2007dynamical, garrahan2009first, jack2010large, chetrite2013nonequilibrium, jack2015effective, jack2020ergodicity, fodor2022irreversibility, agranov2022entropy, agranov2023tricritical, carugno2023delocalization}. In this approach, the ensemble of equilibrium trajectories is biased by conditioning on the value of a non-equilibrium observable such as a time-averaged current. An essentially equivalent (by analogy with equilibrium statistical mechanics) trajectory ensemble can then be constructed by application of an appropriate biasing field that is conjugate to the desired observable. Either way, the bias favours trajectories that in equilibrium would be rare fluctuations, thus driving the system out of equilibrium. One attraction of such ensembles, known variously as \emph{biased}, \emph{tilted} or $s$--ensemble~\cite{garrahan2009first, chetrite2015nonequilibrium, chetrite2015variational, jack2020ergodicity}, is that they are ``minimally biased'' in the sense that they maximize the entropy in trajectory space~\cite{PreGhoLeeDil13} subject to the desired constraint. {Physically, the biasing field allows one to study the properties of rare trajectories that realize, in the undriven dynamics, unusually high or low values of the observable of interest.}
%are favoured and hence, in the long--time limit,  the mean values of the physical observables change with respect to the equilibrium one. In other words, the system is driven out of equilibrium, and a different ensemble is created depending on the value of the imposed field. The formal mathematical construct for this setting is known as \emph{biased}, \emph{tilted} or $s$--ensemble~\cite{chetrite2015nonequilibrium, garrahan2009first, jack2020ergodicity}.

So far, the above two paradigms have largely been considered separately~\footnote{One recent exception comes from studies of the plastic activity under driving oscillatory strain, which exhibits slow power laws suggestive of aging~\cite{ParSasSol22}.}. In particular, studies of driving by trajectory bias have focussed on cases where non-equilibrium steady states are reached, and the dynamical phase transitions between those as the bias is varied~\cite{garrahan2007dynamical, garrahan2009first, jack2010large, chetrite2013nonequilibrium, jack2015effective}. 
{Studies of driven binary Lennard--Jones mixtures in the supercooled regime~\cite{hedges2009dynamic, speck2012first, royall2020dynamical}, for example, have demonstrated %There, it is revealed 
the existence of a transition 
%dynamical first--order phase transition
between an active {(i.e.\ ergodic)} and an inactive  {(i.e.\ nonergodic)} phase, with the physical dynamics being at coexistence between these, but could not address the aging dynamics in the glassy regime at lower temperatures.}
%being the supercooled state the coexistence point. In this work, we focus on rather simplified models with an achievable glassy phase.} 
Here we therefore bring the two paradigms together and study the driven dynamics of aging systems. Our results show the existence of two qualitatively different classes of aging dynamics: {\em fragile} aging that is destroyed by any bias, leading either to an active steady state or an inactive frozen state, and {\em robust} aging that persists even in the presence of bias. As regards our methodology, the \mbox{$s$--ensemble} can generally be studied by analysing the properties of a biased version of the master operator, {i.e.\ of the transition rate matrix}~\cite{derrida1998exact, lebowitz1999gallavotti, garrahan2009first, jack2010large,%
chetrite2013nonequilibrium}; as long as this has a finite spectral gap, the driving leads to a steady state~\cite{jack2010large}. To analyse the driven dynamics of aging systems, we therefore construct an appropriately generalized approach that is applicable also to systems with a vanishing spectral gap.

We consider in this study trap models (TMs), which are paradigmatic descriptions of slow, glassy dynamics. They represent such dynamics as a sequence of jumps between local energy minima or ``traps'' in configuration space, and have played a key role in understanding the physical properties of glasses, see e.g.~\cite{Dyre87, Bouchaud92, denny2003trap, heuer2008exploring, scalliet2021excess, ridout2023building}. What makes them attractive for our purposes is that -- in the absence of driving -- they exhibit well understood aging behaviour~\cite{Bouchaud92, godreche2001statistics, arous2002aging, BerBir11, bertin2013ageing, baity2018activated}. The key signature of this is that correlation functions become explicitly dependent on two times rather than just time differences. Equivalently, one can say that relaxation timescales grow (in principle without bound) with the age of the system, measured from the time of its preparation in the glassy phase.

Mathematically, TMs are defined as continuous time, discrete state Markov chains with random energy landscapes~\cite{MonBou96}. The disorder in the energies (or trap depths) generates a broad distribution of relaxation times characteristic of systems with glassy dynamics~\cite{ArcLanBerBir21, scalliet2021excess}. %At the level of the properties of the master operator, t
This translates into a vanishing spectral gap (in the thermodynamic limit) of the master operator, and in fact the spectral density of the slowest modes generically exhibits a power-law singularity below the glass transition temperature~\cite{margiotta2018spectral, margiotta2019glassy, tapias2020entropic, tapias2022localization}.

\begin{figure}
%\begin{subfigure}[b]{\linewidth} 
  % \includegraphics[width=\linewidth]{FiguresPaper/phase_diagram_bou.pdf}
  \includegraphics[width=\linewidth]{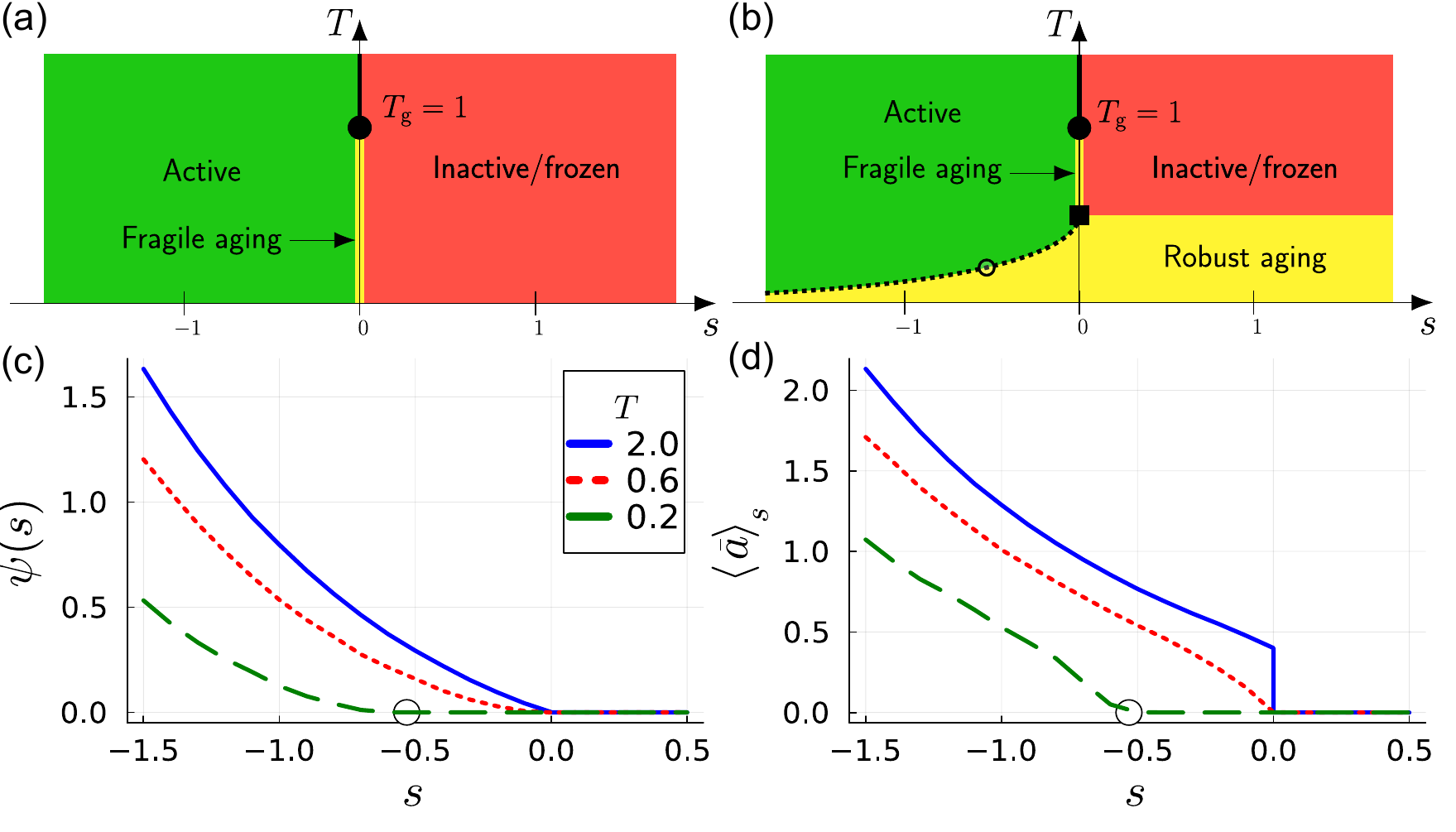}
%\caption{}\label{phases}   
%\end{subfigure}
%\mbox{}\\
%\begin{subfigure}[b]{\linewidth}
%\includegraphics[width=\linewidth]{FiguresPaper/phase_diagramBM.pdf}  
  %\includegraphics[width=\linewidth]{FiguresPaper/phase_diagram_bm.pdf}
%\caption{}\label{phaseBM}
%\end{subfigure}
%\begin{subfigure}[b]{\linewidth}
%\includegraphics[width=\linewidth]{FiguresPaper/energy_activity_barrat_mezard.eps}
%\caption{}\label{phianda}
%\end{subfigure}
\caption{Dynamical $(s, T)$ phase diagrams for (a) Bouchaud and (b) Barrat--M\'ezard trap models;  (c) dynamical free energy  and (d) mean time--averaged activity as a function of the bias parameter $s$ for Barrat--M\'ezard model, for different temperatures $T$ as shown. {In (a) and (b) the solid line on the $T$-axis above $T_{\rm{g}}$ indicates a first--order phase transition. Below $ T_{\rm{g}}$ the transition is continuous and for BM it} is shifted away from $s=0$ to negative $s$ for $T<1/2$, indicating a robust aging phase. {The circles in (c,d) correspond to the circle on the dynamical phase transition line in (b).
% corresponds to those A specific point at the phase boundary is highlighted in figure (b) and replotted in (c) and (d) to illustrate the consistency of the different diagrams.
} 
}
\label{mainplot}
\end{figure}

%Application of the $s$--ensemble to KCMs has revealed a dynamical first-order phase transition between an active and an inactive phase that coexist for any finite temperature in the unbiased system ($s=0$). The active phase is characterized by an extensive (in system size and observation time) number of kinetic transitions and therefore a finite time--averaged activity, whereas the inactive phase is characterized by a subextensive number of transitions for long observation times that lead to zero activity~\cite{garrahan2007dynamical, garrahan2009first}. The emergent picture from this analysis turns out to be relevant for the description of the supercooled liquid state as a coexistence state in trajectory space~\cite{chandler2010dynamics, royall2020dynamical} .

In this work, we investigate the effects of driving on the aging dynamics of the two most widely studied mean--field TMs as defined by Bouchaud~\cite{Bouchaud92} and Barrat and M\'ezard (BM)~\cite{barrat1995phase}. The driving is chosen to bias trajectories towards high or low activity, defined as the number of jumps from one trap to another. A pictorial representation of our main result is given in Fig.~\ref{mainplot}a \& b, where we show the phase diagrams in the plane of bias parameter $s$ and temperature $T$. These diagrams indicate that, as analysed in detail below, the BM trap model has a robust aging phase at low $T$ while aging in the Bouchaud model is always fragile, thus revealing two qualitatively distinct patterns in the response of aging to driving. 
% together with representative curves of the dynamical free energy and the mean time--averaged activity for BM model (quantities defined few paragraphs below).

We outline briefly the commonalities and differences between the TMs we consider, before introducing the driving by activity bias. Both models models assume a disordered landscape of traps, with depths $E>0$ drawn from an exponential distribution $\rho(E) ={\rm{e}}^{-E}$ whose width sets our units of energy. This leads to the emergence of a glass transition at temperature $T_{\rm{g}}=1$; above this, there is an well--defined equilibrium phase with an associated  Boltzmann distribution. Below $T_{\rm{g}}$ there is no steady--state solution for the master equation governing the dynamics (eq.~\eqref{mastertm} below), resulting in slow, glassy relaxation and aging.
%distribution becomes non--normalizable and therefore the partition function diverges in the thermodynamic limit.
%Aging and glassy dynamical properties are exhibited in this regime.
The two models differ in the choice of transition rates between configurations (or traps): for the Bouchaud TM they are of Arrhenius form  while the BM model posits Glauber rates:
% whereas for BM, Glauber--like, explicitly they are
\begin{align}
    W_{\rm{Bouchaud}}(E \to E') &= {\rm{e}}^{-E/T}   \label{boutr}  \\
    W_{\rm{BM}}(E \to E') &= 1/[1+ {\rm{e}}^{-(E' - E)/T}] 
    \label{bmtr}
\end{align}
Assuming mean field connectivity where any trap can be reached from any other, the (undriven) dynamics of a TM in the thermodynamic limit of a large number traps is then described by a master equation for the probability $P(E, \tau)$ of being in a trap of depth $E$ at time $\tau$:
\begin{align}
\partial_\tau P(E,\tau) &= -  r(E) P(E,\tau) + \notag  \\
    &\rho(E) \int_0^\infty dE' \, W(E' \to E) P(E',\tau) 
    \label{mastertm}
\end{align}
with $r(E) = \int dE' \,W(E \to E') \rho(E') $ the escape rate from a trap of depth $E$. This is
simply ${\rm{e}}^{-E/T}$ for Bouchaud, and $\propto {\rm{e}}^{-E}$ for sufficiently deep traps in BM~\cite{SM}. {Formally, the master operator is the linear operator that produces the r.h.s.\ of equation~\eqref{mastertm} when acting on $P(E, \tau)$.}
\iffalse
 given by
% $r(E)$ the escape rate of the corresponding (unbiased) trap model, namely
  \begin{subnumcases}{r(E)=}
    {\rm{e}}^{-E/T} \qquad \qquad &{\rm{Bouchaud}}, \label{ratebou} \\
   \frac{\pi T}{\sin(\pi T)} \,e^{-E} \quad &{\rm{Barrat--M\'ezard}}. \label{ratebm}
  \end{subnumcases}
For the BM case, the above is the limiting form for deep traps, which determines the aging dynamics; the full expression can be written as a hypergeometric function.
% F(1,T,1+T, -{\rm{e}}^{E/T}) 
\fi

To implement the driving we bias trajectories of the above two TMs according to their activity~\cite{merolle2005space, jack2006space, garrahan2007dynamical, hedges2009dynamic, jack2010large, chetrite2013nonequilibrium, jack2020ergodicity} $\mathcal{A}$, defined as the total number of jumps between traps along a trajectory of duration $t$ {
(and distinct from the notion of activity in active matter, where the term normally refers to the strength of motility)%
%this term is conceptually different from the one used in the context of active matter,  where activity typically refers to the strength of motility)
}. We write $\bar{a}=\mathcal{A}/t$ for the time-averaged activity. Introducing a conjugate bias field $s$, %conjugate to the activity, 
the driven trajectory ensemble is then defined by reweighting each trajectory with ${\rm{e}}^{-s\mathcal{A}}={\rm{e}}^{-st\bar{a}}$, with positive $s$ favouring smaller $\mathcal{A}$. The appropriate normalization constant is the dynamical partition function
%\begin{align}
$    Z(s,t) =  \langle {\rm{e}}^{-s t \bar{a}} \rangle_0
$,
%    \label{parti}
%\end{align}
%where 
the average being over the trajectory distribution of the undriven dyamics ($s = 0$). Associated with this is a 
{\em dynamical free energy}
% associated with the distribution of $\bar{a}_t$ is then defined as
\begin{align}
    \psi(s) = \lim_{t \to \infty} \frac{1}{t} \ln \langle {\rm{e}}^{-s t \bar{a}} \rangle_0
    \label{dynf}
\end{align}
that can alternatively be seen as the scaled cumulant generating function of $\bar{a}$. By direct analogy with equilibrium thermodynamics, 
%Following the thermodynamic construction, it is common to express the dynamical free energy in terms of a partition function, which would correspond to the quantity in angular brackets in equation~\eqref{dynf}, this is
%\begin{align}
%    Z(s,t) :=  \langle {\rm{e}}^{-s t \bar{a}_t} \rangle_0
%    \label{parti}
%\end{align}
%From equation~\eqref{dynf} one obtains that %the mean time--averaged activity under the biasing field in the long time limit is  
derivatives w.r.t.\ the field $s$ give averages over the biased ensemble: %of the observable: leaving the limit $t\to\infty$ implicit one has
\begin{align}
  -\psi'(s) = -\frac{1}{t} \frac{\partial}{\partial s}  \ln Z(s,t) = \frac{\langle \bar{a}\,{\rm{e}}^{-st\bar{a}} \rangle_0}{Z(s,t)} = \langle \bar{a} \rangle_s
  \label{psi_prime}
\end{align}
where we have left the limit $t\to\infty$ implicit and the last equality defines the mean time-averaged activity in the driven system. 
%which is defined as the mean value of the time--averaged activity under the biasing field $s$.
%In agreement with the definition of the driven trajectory ensemble, the penultimate expression shows that %An implication from this is that 
%positive (negative) bias $s$ favours smaller (larger) values of the activity compared to those in the unbiased system.
Higher-order cumulants can be obtained similarly, as well as -- via a Legendre transform and under appropriate assumptions --
%one can further
% the assumption that $\bar{a}$ obeys a large deviation principle, one can 
%extract from $\psi(s)$ %not just the mean activity but 
%via a Legendre transform %the dynamical free energy and 
the rate function for the large deviations of $\bar{a}$~\cite{chetrite2013nonequilibrium, jack2020ergodicity}.

The dynamical free energy $\psi(s)$ can be calculated as the largest eigenvalue of a biased master operator $\WW(\g)$~\cite{derrida1998exact, lebowitz1999gallavotti}. The latter retains the diagonal elements of the original master operator, i.e.\ the escape rates $r(E)$,  but 
changes the off--diagonal elements, i.e.\ the transition rates, into 
\begin{align}
    W(E \to E', s) = W(E \to E') {\rm{e}}^{-s} 
    \label{biased}
\end{align}
This comes from the fact that the weight of any trajectory in the driven ensemble is multiplied by ${\rm{e}}^{-s}$ for each jump. For the Bouchaud TM, the resulting eigenvalue problem for $\psi(s)$ can be reduced to a single self-consistent equation~\cite{SM}. 
%Taking then the derivative w.r.t.\ $s$ according 
Determining the mean time-averaged activity from~\eqref{psi_prime} we find that  $\langle\bar{a}\rangle_s$ is positive for $s<0$ at all $T$, while it vanishes for $s>0$ (where also $\psi(s)=0$). In the first regime we therefore have an {\em active} steady state while for $s>0$ the system is {\em inactive}, see Fig.~\ref{mainplot}a.
%
%Let us now elaborate more on our results. The phase diagram in Fig.~\ref{mainplot} reveals that the physical equilibrium phase ($T > T_g$ at $s=0$) corresponds to a phase coexistence (first order phase transition) between an active and an inactive (or frozen) phase
In these broad features the phase diagram resembles previous results for kinetically constrained models, with a dynamical phase transition at $s=0$. 
There, however, the transition is always first order~\cite{garrahan2007dynamical, garrahan2009first}. For the Bouchaud model, this scenario applies only above the glass transition, $T>1$, while for lower temperatures we find that the transition is second order. This in fact makes sense: at $s=0$ for $T<1$ the system {\em ages} and its mean time-averaged activity $\langle\bar{a}\rangle_0$ vanishes for $t\to\infty$. The {\em driven} activities $\langle\bar{a}\rangle_s$ for $s<0$ then extrapolate to this zero value as $s\to 0$,
%from negative values, 
making the transition continuous; we find explicitly the scaling $\langle\bar{a}\rangle_s \sim |s|^{\beta-1}$ (with $\beta=1/T$ as usual)~\cite{SM}.
% or equivalently the total activity grow
%~\footnote{The free energy vanishes for $s > 0$ and is given by equation~\eqref{free1} around the transition for $s < 0$ for Bouchaud model. For BM, numerics suggest same scaling with a different prefactor}. The active phase is characterized by a positive dynamical free energy and a finite mean time--averaged activity. On the other hand, the inactive phase, exhibits a vanishing of both the free energy and the mean time--averaged activity as a result of a finite number of jumps (in the $t \to \infty$ limit). 
The point $(s = 0, T = 1)$, i.e.\ the glass transition of the undriven system, thus becomes a tricritical point in the $(s,T)$-phase diagram, at which the order of the transition changes from first to second {(we refer to Refs.~\cite{agranov2022entropy, agranov2023tricritical} for other recent examples of tricritical behavior in the context of dynamical phase transitions.)}

For the BM model we find
% the eigenvalue calculation for $\psi(s)$ is more involved but 
for $T>1/2$ %we nonetheless find~\cite{SM} 
a qualitatively similar dynamical phase diagram~\cite{SM}, with a transition from an active to an inactive phase at $s=0$ that is first order for $T>1$ and second order for $1/2<T<1$.
% a the same phase boundary is present for $T > T_g/2$. However, below $T_g/2$ it shifts towards negative values of the bias, eventually diverging at zero temperature. We stress that for both models below $T_g$, the phase boundary does not correspond to a dynamical first--order transition as the first derivative of the dynamical free energy is continuous across the boundary.% Indeed, for the Bouchaud model one can show (see supplemental material) that the transition is second order.
This agreement between the two models for $T>1/2$ %on the phase diagram for $T> T_g/2$ between both models is not surprising, 
is physically reasonable because 
%, as has been discussed elsewhere, 
in this regime the slow dynamics in the BM model can be shown to arise from activation across {\it effective} energy barriers~\cite{bertin2003cross, sollich2006trap, cammarota2015spontaneous}, mirroring the explicitly activated dynamics~\eqref{boutr} in the Bouchaud TM.
For lower temperatures, $T<1/2$, the BM dynamical phase diagram becomes qualitatively different, with the phase boundary moving towards negative $s$ (%black dashed line for $T < 1/2$ in 
Fig.~\ref{mainplot}b). Using an appropriate scaling assumption for the eigenvector~\cite{SM} we can determine its location in closed form as
\begin{align}
   s^* = \ln (\sin(\pi T)) 
    \label{phbound}
\end{align}
The absence of a transition at $s=0$ already hints at the fact that low temperature aging in the BM model is robust to driving, a point we will substantiate below. The shift of the phase boundary away from $s=0$ also
reflects a change in character of (undriven) BM aging: for $T<1/2$ the slow dynamics is governed by entropic barriers, i.e.\ by the scarcity of ever deeper traps%
% as the system descends in its energy landscape over time%. entropically driven, meaning that the evolution depends on the existence of directions towards deeper traps, that become scarcer as time passes
~\cite{barrat1995phase, bertin2003cross, sollich2006trap}.  
% We elaborate more on this regime later below.

\begin{figure}
%   \begin{subfigure}[b]{.9\linewidth}
% \includegraphics[width=\linewidth]{FiguresPaper/timeacti1.pdf}
% \caption{}\label{tacti1}
% \end{subfigure}
% \begin{subfigure}[b]{.9\linewidth}
% \includegraphics[width=\linewidth]{FiguresPaper/timeacti2.pdf}
% \caption{}\label{tacti2}
% \end{subfigure}
  \includegraphics[width=\linewidth]{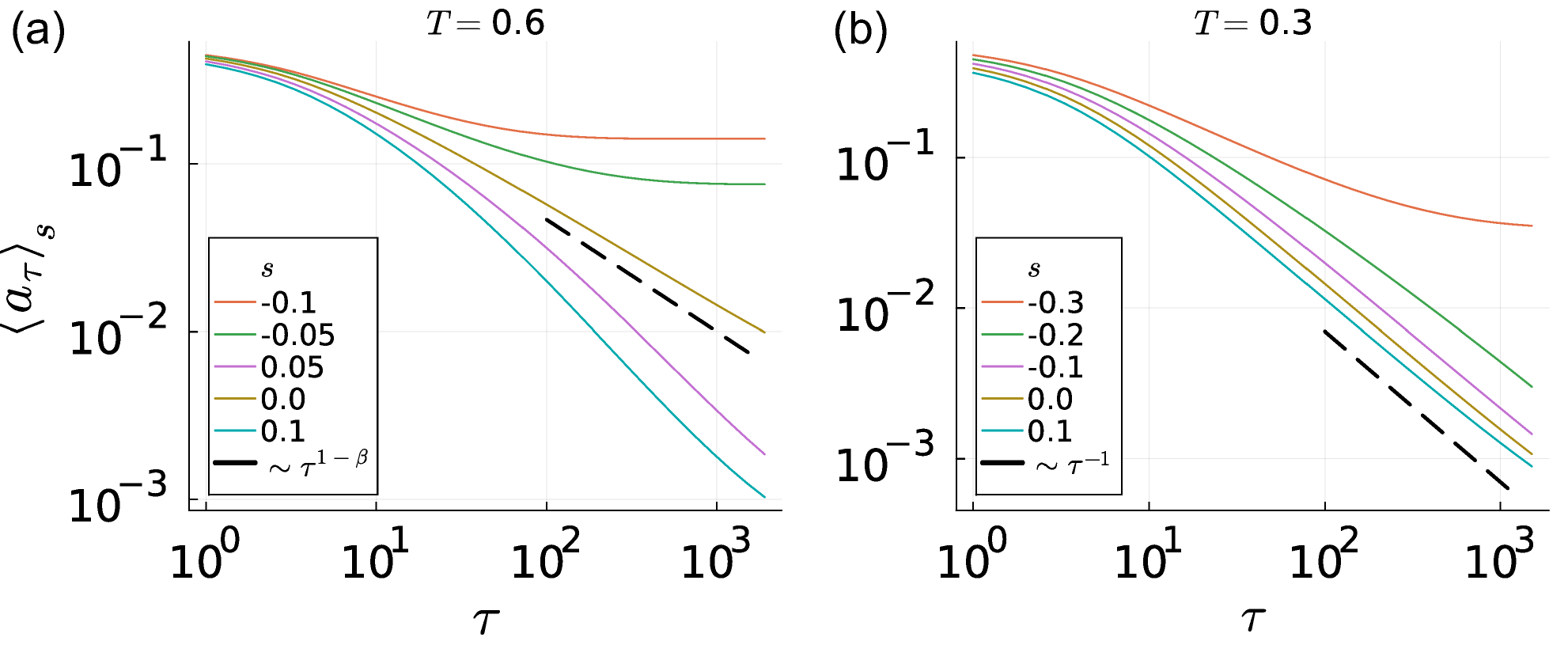}
  \caption{Mean time--dependent activity for two temperatures and different driving parameters $s$ as shown, for BM trap model. Dashed lines: power law scaling for the undriven case ($s=0$). %Total trajectory length $t = 10^4$.
  }
   \label{timeactivi}
 \end{figure}

\begin{figure}
%\centering
%\begin{subfigure}[b]{.9\linewidth}
%\includegraphics[width=\linewidth]{FiguresPaper/dist_positive_bias_bm_highT.pdf}
% \caption{}\label{distribm1}
% \end{subfigure}
% \begin{subfigure}[b]{.9\linewidth}
% \includegraphics[width=\linewidth]{FiguresPaper/dist_positive_bias_bm_lowT.pdf}
% \caption{}\label{distribm}
% \end{subfigure}
  \includegraphics[width=\linewidth]{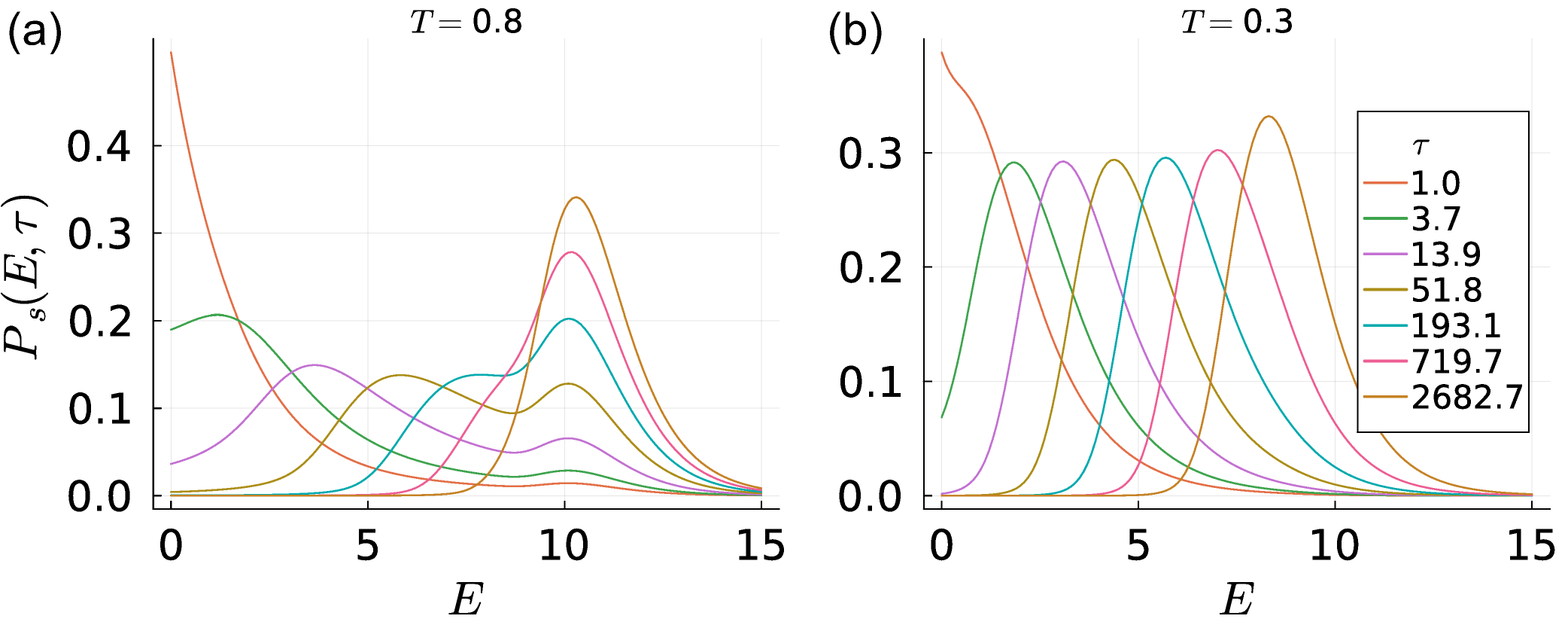}
\caption{Time evolution of trap depth distribution for BM trap model with fixed driving parameter $s=0.25$ at temperatures %$s$ and total time $s = 0.25, t = 10^4$ and two different temperatures (
(a) above and (b) below $1/2$. Note in (b) the continual shift of the distribution to larger trap depths indicating robust aging. % even in the presence of driving by trajectory bias. %Total trajectory length $t=10^4$.
}
\label{distributions}
\end{figure}

While the above free energy calculations can distinguish active and inactive dynamical phases, we need more precise tools to study whether the driven systems still show {\em aging}. In undriven TMs, the aging regime exhibits %can be identified by
a vanishing time-averaged activity, i.e.\ $\langle\bar{a}\rangle=\langle\mathcal{A}\rangle/t\to 0$ for $t\to\infty$, but as the system keeps evolving its {\em total} activity $\langle \mathcal{A}\rangle$ still diverges (sublinearly) with trajectory length $t$. We therefore now ask {whether} this remains the case in the inactive ($\langle \bar{a}\rangle \to 0$) regions of our phase diagrams. This would be straightforward for the undriven dynamics: one writes $\mathcal{A}=\int_0^td\tau\,a_\tau$ where $a_\tau$ is the time-dependent activity or rate of jumps. Using that the mean of the latter is the mean escape rate
\begin{equation}
\langle a_\tau\rangle = 
%\langle r_\tau\rangle = 
\int dE\,P(E,\tau)r(E) 
\end{equation}
one then only needs to solve the Master equation~\eqref{mastertm} to determine $\langle a_\tau\rangle $. This reasoning does not work, however, for the auxiliary master operator~\eqref{biased} as this does not represent a physical, probability-conserving dynamics. Consequently, we develop a formalism to represent the driven dynamics as an equivalent ``auxiliary'' Markov process, extending existing work%
%In order to understand better the physics at different points of the phase diagram and to stress the differences between both models we introduce the formalism to analyze the time evolution of the driven system
~\cite{jack2010large, chetrite2013nonequilibrium, chetrite2015nonequilibrium, jack2020ergodicity} to %, in a form suitable for 
aging systems. % without a spectral gap. 
% In the presence of a bias field, the probability of any path is naturally changed with respect to the unbiased system. To derive an expression that reflects this change, we introduce a biased master operator $\WW(\g)$ that preserves the diagonal elements of the original master operator but changes the off--diagonal, according to
%\begin{align}
%    W(E \to E', s) = W(E \to E') {\rm{e}}^{-s}
%    \label{biased}
%\end{align}
%The top eigenvalue of this operator corresponds to the dynamical free energy $\psi(s)$.  However, this operator by itself does not conserve probability and therefore the solution of the associated master equation (denoted as the forward master equation) is not completely meaningful by itself.
%It is possible, anyways, to construct a proper (time--dependent) stochastic master operator, $\WW^{\rm{aux}}(\tau)$  that preserves the probability of the biased trajectories. 
The appropriate auxiliary master operator (see also remarks for diffusion processes in~\cite{chetrite2015nonequilibrium, chetrite2015variational}) has transition rates~\cite{SM}
% of the unbiased system as
\begin{align}
    W^{\rm{aux}}(E \to E', \tau) = W(E \to E')\, {\rm{e}}^{-s} \,\frac{q(E', \tau)}{q(E,\tau)}
    \label{auxi}
\end{align}
with $q(E, \tau)$ the % time--dependent function that  solves 
solution of the backward master equation for $\WW(\g)$,
\begin{align}
  -   % \frac{\partial}
  {\partial_\tau} q(E, \tau) &= - r(E) q(E, \tau)\\
                                                 &+ {\rm{e}}^{-s}  \int dE'  q(E',\tau) W(E \to E') \rho(E')
\notag 
\label{backman}
\end{align}
and final time boundary condition $q(E, t) = 1$. The corresponding auxiliary escape rates %, on the other hand, 
are given by
\begin{align}
r^{\rm aux}_\tau(E) = r(E)-\partial_\tau \ln(q(E,\tau)) 
\end{align}
%Within this construction, the finite mean time--averaged activity corresponds to the mean auxiliary escape rate~\footnote{See suplemental material for a full detailed derivation}. This quantity by itself is computed as
Within this probability-conserving Markov dynamics we {\em can} now compute the desired time-dependent activity as the mean escape rate:
\begin{align}
  \langle a_\tau \rangle_s &=  \int dE\,P_s(E, \tau) [r(E)- \partial_\tau \ln(q(E, \tau) ]
%\, , \\
%  &= \langle r_\tau \rangle_s -  \langle \partial_\tau \ln(q(E, \tau)   \rangle_s
    \label{meana}
\end{align}
It then remains to find the distribution of trap depths in the driven systems, 
%Coming back to eq.~\eqref{meana}, it is averaged over 
$P_s(E, \tau)$, that enters into this average. This has the form
%is the probability distribution associated with the auxiliary master operator, computed generically as
\begin{align}
  P_s(E, \tau) = \frac{q(E, \tau) P_{\rm{f}}(E, \tau)}{Z(\g, t)} 
  \label{compsolution}
\end{align}
where $P_{\rm{f}}(E, \tau)$ is the solution of the forward master equation for $\WW(s)$.
%once this is known, the partition function can be found using %From eq.~\eqref{compsolution} and 
%the final time boundary condition for 
%%the backwards factor 
%$q$
%%, the partition function can also be expressed 
%as
%\begin{align}
%  Z(s,t) = \int dE\,\Pf(E, t)
%  \label{parte}
%\end{align}
In the simple case where $\WW(\g)$ has a spectral gap~\cite{jack2010large}, $q(E,\tau)\propto {\rm{e}}^{\psi(s)(t-\tau)}$ while $P_s(E,\tau)$ reaches a steady state for large $\tau$~\footnote{Both statements apply away from the ends of a long trajectory, i.e.\ once $\tau$ and $t-\tau$ are significantly larger than the inverse spectral gap.} and one retrieves the known result $r^{\rm aux}(E)=r(E)+\psi(s)$~\cite{evans2004rules, evans2004detailed} and a time-independent nonzero activity 
%For large observation times $t \to \infty$ and long times $\tau \gg 1$, it must happen that
%\begin{align}
%      \langle a_\tau \rangle_s \to     \langle \bar{a} \rangle_s 
%\end{align}
%In the active phase, a stationary state is reached, which is implied by the existence of a spectral gap in the biased master operator~\cite{jack2010large}. In such a case, equation~\eqref{meana} simplifies to 
\begin{align}
  \langle a_\tau \rangle_s \to \langle \bar{a}\rangle = \langle r \rangle_{ s} + \psi(s) 
  \label{meanac}
\end{align}
%and it yields a finite value in the long--time limit (see Fig.~\ref{phianda}(right)). 
This approach of the activity to a steady state plateau can be seen for the most negative $s$-values in Fig.~\ref{timeactivi}, consistent with an active phase. In the inactive phase, we observe in the BM model with $T<1/2$ (Fig.~\ref{timeactivi}b) that the activity $\langle a_\tau\rangle_s$ decays like in the undriven case $s=0$, as $\sim 1/\tau$~\footnote{This can be confirmed analytically at $T=0$, for any bias $s$~\cite{SM}.}: the system continues to age in the presence of driving, its total activity $\langle\mathcal{A}\rangle_s$ growing as $\ln t$. This is what we mean by {\em robust aging}, and it is confirmed by Fig.~\ref{distributions}b: the distribution of trap depths continues to shift (logarithmically) to larger values with increasing $\tau$, just as in undriven aging.

The effect of driving on aging in the BM model is rather different for $T>1/2$: while for $s=0$ one has $\langle a_\tau\rangle_0 \sim \tau^{1-\beta}$~\cite{bertin2003cross} and so a divergent total activity $\langle \mathcal{A}\rangle_0 \sim t^{2-\beta}$ demonstrating aging, any $s>0$ leads to a faster activity decay $\langle a_\tau \rangle_s  \sim \tau^{-(\alpha+1)}$ with $\alpha>0$ and thus a {\em finite total} activity $\langle \mathcal{A}\rangle_s$. This means the inactive phase is {\em frozen} here, with the system coming to rest after a finite number of jumps. Correspondingly, $P_s(E, \tau)$ in Fig.~\ref{distributions}a shows a rather rapid transfer of probability with increasing $\tau$ into traps that are sufficiently deep for the system not to make further jumps in the remainder of the trajectory. Overall, the aging in the undriven system ($s=0$) is {\em fragile}: any positive drive freezes the system, while any negative $s$ produces an active steady state. The Bouchaud model behaves similarly throughout its glass phase, with an aging activity decay $\langle  a_\tau\rangle_0 \sim \tau^{T-1}$ in the undriven case $s=0$~\cite{Bouchaud92} that changes to $\langle a_\tau \rangle_s \sim \tau^{-(T+1)}$~\cite{SM} and hence a frozen, finite total activity phase, for any $s>0$.

Summarizing, we have shown that the interplay of aging and driving by trajectory bias leads to significant new insights into aging dynamics. In the trap models we studied, which in the absence of driving exhibit aging below their glass transition temperature, this aging dynamics can respond in two fundamentally different ways to driving. In the BM model for $T>1/2$ and in the Bouchaud model, aging is fragile: any drive towards high activity ($s<0$) produces an active steady state, while driving towards low activity leads to frozen dynamics arresting after a finite number of jumps. For $T<1/2$ in the BM model, on the other hand, aging is robust, remaining qualitatively unchanged by moderate driving towards either higher or lower activity. We demonstrated these differences using a general analysis framework for Markovian systems driven by trajectory bias, which is complete in the sense that it can be deployed whether or not the underlying undriven system exhibits aging. What is fascinating in particular is that the response to driving detects, without additional physical input via e.g.\ specifically designed coarse-graining~\cite{cammarota2015spontaneous, cammarota2018numerical, baity2018activated}, the qualitative differences between aging dynamics governed by energetic barriers {(as in the Bouchaud model and the BM model with $T >1/2$)} -- which we found to be fragile -- and by entropic effects. The resulting robust aging behavior that we found in the low--temperature BM model may have broader implications for the physics of supercooled liquids, bearing in mind a recent proposal that entropic relaxation is dominant in this {regime}~\cite{baity2021revisiting}.  {To shed further light on the nature of the aging phases one could convert our results into statements about activity fluctuations in the unbiased dynamics.
% is given by putation of the corresponding rate functions. 
We do this in~\cite{SM} for the Bouchaud model; an analogous calculation for the BM model, yielding additional signatures of robust aging, remains an exciting open challenge.}  
Generally, the response of aging systems to driving that we have studied here, by bringing together the two key non-equilibrium paradigms, provides new analysis tools for non-equilibrium dynamics and could eventually lead to a classification of aging into distinct universality classes.

{\bf Acknowledgements}: FAL acknowledges financial support through a scholarship from Conacyt (Mexico). {Simulations were run on the GoeGrid cluster at the University of Göttingen, which is supported by the Deutsche Forschungsgemeinschaft (project IDs 436382789; 493420525).}

\bibliography{cavi}

%\putbib
%\end{bibunit}

\pagebreak
%\newpage
\widetext

\begin{center}
  \textbf{\large Bringing together two paradigms of non-equilibrium:\\ Fragile versus robust aging in driven {glassy} systems
 \\Supplementary Material}
  \end{center}
% \\[.2cm]
%   Diego Tapias,$^{1,*}$ Charles Marteau,$^{2,3}$ , Fabián Aguirre-López$^{2,4,5}$ and Peter Sollich$^{1,2, \dag}$\\[.1cm]
%   {\itshape ${}^1$ Institute for Theoretical Physics, Georg-August-Universit\"at G\"ottingen, 37077 G\"ottingen, Germany\\
%     ${}^2$ Department of Mathematics, King's College London, London WC2R 2LS, UK \\
%     ${}^3$ Department of Physics and Astronomy, University of British Columbia, Vancouver V6T 1Z1, Canada \\
%   ${}^4$ Chair of Econophysics and Complex Systems, École polytechnique, 
% 91128 Palaiseau Cedex, France \\
%   ${}^5$  LadHyX UMR CNRS 7646, École polytechnique, 91128 Palaiseau Cedex, France \\} 
% ${}^{*}$Electronic address: diego.tapias@theorie.physik.uni-goettingen.de \\
% ${}^\dag$Electronic address: peter.sollich@uni-goettingen.de \\
% (Dated: \today)\\[1cm]
% \end{center}

\setcounter{equation}{0}
\setcounter{figure}{0}
\setcounter{table}{0}
\renewcommand{\theequation}{S\arabic{equation}}
\renewcommand{\thefigure}{S\arabic{figure}}
%\renewcommand{\bibnumfmt}[1]{[S#1]}
%\renewcommand{\citenumfont}[1]{S#1}

% \\[.2cm]
%   Diego Tapias,$^{1,*}$ Charles Marteau,$^{2,3}$ , Fabián Aguirre-López$^{2,4,5}$ and Peter Sollich$^{1,2, \dag}$\\[.1cm]
%   {\itshape ${}^1$ Institute for Theoretical Physics, Georg-August-Universit\"at G\"ottingen, 37077 G\"ottingen, Germany\\
%     ${}^2$ Department of Mathematics, King's College London, London WC2R 2LS, UK \\
%     ${}^3$ Department of Physics and Astronomy, University of British Columbia, Vancouver V6T 1Z1, Canada \\
%   ${}^4$ Chair of Econophysics and Complex Systems, École polytechnique, 
% 91128 Palaiseau Cedex, France \\
%   ${}^5$  LadHyX UMR CNRS 7646, École polytechnique, 91128 Palaiseau Cedex, France \\} 
% ${}^{*}$Electronic address: diego.tapias@theorie.physik.uni-goettingen.de \\
% ${}^\dag$Electronic address: peter.sollich@uni-goettingen.de \\
% (Dated: \today)\\[1cm]
% \end{center}

\setcounter{equation}{0}
\setcounter{figure}{0}
\setcounter{table}{0}
\renewcommand{\theequation}{S\arabic{equation}}
\renewcommand{\thefigure}{S\arabic{figure}}
%\renewcommand{\bibnumfmt}[1]{[S#1]}
%\renewcommand{\citenumfont}[1]{S#1}

%\pagebreak
%\widetext

%\begin{bibunit}
\subsection{Biased ensembles of trajectories}
\label{biasedensemble}

In this section we present the formalism of biased ensembles of trajectories applied to a general Markov chain, with the aim of deriving, and providing intuition for, the main equations presented in the text. We follow closely the setup in Ref.~\cite{jack2010large}. We present the formalism for a discrete configuration space first. The appropriate modifications for the thermodynamic limit with its continuous trap depth distributions is given in a later section.
% we generalise some of the equations to a continuous space.

Consider a Markov chain on a set of configurations labelled by $\C$. The system makes stochastic transitions between configurations with transition rates $W(\Cp\to \C)$.  The probability $P(\C, \tau)$ of finding the system in configuration $\C$ at
time $\tau$ then evolves according to the master equation 
\begin{equation}
  \frac{\partial}{\partial \tau} P(\C, \tau) = -r(\C) P(\C,\tau) + \sum_{\Cp\neq \C} W(\Cp\to \C) P(\Cp,\tau)
  \label{master}
\end{equation}
with
$r(\C)=\sum_{\Cp\neq \C} W(\C\to\Cp)$
the escape rate from configuration $\C$. We define the vector $|P( \tau)\rangle
=\sum_{\C} p(\C, \tau) |\C\rangle$ (we use bra-ket notation here; $|\C\rangle$ can simply be viewed as the Euclidean unit vector in direction $\C$) and introduce the compact notation $
%{\partial \tau}
\partial_\tau |P( \tau)\rangle=\WW|P( \tau)\rangle$ where the elements of the master operator $\WW$ are $\langle \C|\WW|\Cp\rangle
=W(\Cp\to\C)( 1- \delta_{\C,\Cp}) - \delta_{\C,\Cp} r(\C)$. %The master equation~\eqref{master} is the description of a generic unbiased system.

We consider current-like trajectory observables of the form
\begin{align}
\mathcal{A}
%\bar{a} 
= %\frac{1}{t} 
\sum_{{\rm{jumps}} \, \C\to \C'} \alpha(\C, \C')
% =: \frac{ \mathcal{A}_t}{t}
\label{activ}
\end{align}
where $\alpha(\C,\C')=1$ (for $\C\neq \C'$) gives the activity but other choices can be used to capture currents of particles or entropy etc. We are interested in trajectory ensembles that force the average $\langle \mathcal{A}\rangle$ to be far
from its average value  $\langle\mathcal{A}\rangle_0$. This is most conveniently done not via a hard constraint but with a bias %towards atypical values of $\bar{a}_t$
in the probability assigned to each trajectory $\pi$:
\begin{equation}
P[\pi,\g] = \frac{P[\pi,0]  
  \exp\left( -\g \mathcal{A} \right)}{ Z(\g,t)}
\label{equ:s-ens}  
\end{equation}
After discretizing the total trajectory length $t$ into small steps $\dt$, a trajectory $\pi$ is represented as a sequence of configurations $\{\C_\tau\}$ for $\tau=0,\dt,2\dt,\ldots, t$. In terms of the original master operator $\WW$, the biased probability of a trajectory 
%($\pi=\{\C_\tau\}$ in equation~\eqref{equ:s-ens}) 
can then be written as
\begin{align}
P[\pi,\g] &= 
 \frac{\left[\prod_{\tau=0}^{t-\dt} \langle \C_{\tau+\dt}| (\mident + \dt \WW)|\C_\tau\rangle \eee^{-\g\alpha(\C_\tau,\C_{\tau+\dt})} 
\right]}{Z(\g,t)} \times p(\C_0,0) 
\label{equ:path_prob}
\end{align}
with a generic initial distribution $|P(0)\rangle = \sum_\C p(\C,0)|\C\rangle$. Here we use the convention $\alpha(\C,\C')=0$ for $\C=\C'$, so that the reweighting factor in eq.~\eqref{equ:path_prob} contributes only when the configuration changes, as defined in~\eqref{activ}. Because of this, the factors in the product can be seen  as matrix elements of a biased master operator defined as (see also~\cite{lecomte2007thermodynamic, garrahan2009first})
\begin{align}
\langle \C | \WW(\g) | \Cp\rangle &= 
\langle \C | \WW | \Cp \rangle \eee^{-\g \alpha(\Cp,\C)}  = 
\left\{ \begin{array}{ll} 
W(\Cp\to\C) \eee^{-\g \alpha(\C',\C)}, \quad & \C\neq\Cp, \\
-r(\C), & \C=\Cp . \end{array} \right.
\label{equ:WA}
\end{align}
Summing equation~\eqref{equ:path_prob} over all possible trajectories gives the normalization factor or partition function
\be 
Z(\g,t) = \langle e| (\mident + \dt \WW(\g))^{t/\dt} |P(0)\rangle 
\to \langle e| \eee^{\WW(\g)t} |P(0)\rangle 
\label{equ:zmat}
\ee
where $\proj=\sum_\C \langle \C|$ is a projection state. 

If the original stochastic model has a finite state space and is irreducible, one sees from the spectral decomposition of the biased master operator
\begin{align}
  \WW(s)=\sum_\alpha |V_\alpha\rangle \lambda_\alpha \langle U_\alpha|
\label{spectral}
\end{align}
that the dynamical free energy (as introduced in eq.~(4) in the manuscript) has a long--time limit~\cite{lebowitz1999gallavotti, lecomte2007thermodynamic, jack2010large, jack2020ergodicity} and is explicitly given by
\begin{equation}
\psi(\g) = \lim_{t\to\infty} \frac{1}{t} \ln Z(s,t)  = \max_\alpha \lambda_\alpha .
\label{equ:res-psi}
\end{equation}
Thus the dynamical free energy is the largest eigenvalue of $\WW(\g)$, which dominates the exponential in (\ref{equ:zmat}) for large $t$. If
$\WW(\g)$ has a spectral gap between 
%its set by the inverse of the gap 
between its largest and second largest eigenvalue, then the biased dynamics approaches a steady state away from the ends ($\tau$ close to $0$ or $t$) of a long enough trajectory. The extent of the temporal boundary regions is given by the inverse spectral gap. %The argument given for equation~\eqref{equ:res-psi} applies generically  in the limit of inifinitely many states to biased operators with non--zero spectral gap.

The systems that we study in the manuscript are interesting because they do not necessarily exhibit a spectral gap. As a matter of fact, in the aging regime the associated relaxation timescales grow without bound (see Refs.~\cite{margiotta2018spectral, tapias2020entropic} for the spectral distribution of the relaxation rates in Bouchaud and BM models, respectively). Then even the unbiased master operator $\WW$ has eigenvalues arbitrarily close to the largest eigenvalue of zero, i.e.\ no gap in the spectrum. In order to analyse these systems and in particular the time-dependence along their trajectories, one needs to introduce a time--dependent \emph{auxiliary master operator}; this construction is described below.

\subsection{Construction of time-dependent auxiliary Markov operator}

The need for this construction arises from the fact that the biased path probability (\ref{equ:path_prob}) is not in the usual form for stochastic Markov dynamics. This is because the corresponding biased master operator does not in general conserve probability, i.e.\ $\proj \WW(\g) \neq  0$. However a multiplicative (``Doob") transform of $\WW(\g)$ can be used to bring it back into the standard form, even for systems without a spectral gap that can exhibit aging. This was shown by Chetrite and Touchette~\cite{chetrite2013nonequilibrium, chetrite2015nonequilibrium} in rather abstract terms for diffusion processes; we make it concrete here for Markov jump processes on discrete state spaces.

One starts by introducing multiplicative factors $q(\C,\tau)$ into the path probability (\ref{equ:path_prob}): 
\begin{align}
  P[\pi,\g] =  \frac{q(\C_0,0) p(\C_0,0)}{Z(\g,t)} \left[\prod_{\tau=0}^{t-\dt} 
q(\C_{\tau+\dt},\tau+\dt)   \langle \C_{\tau+\dt}| (\mident + \dt \WW(\g)) |\C_\tau\rangle q^{-1}(\C_\tau,\tau)\right]
\label{equ:path_prob_multiplicative}
\end{align}
These factors all cancel except for the one at the final time, which therefore has to be $q(\C,t)=1$. The other factors should be chosen so that for every timestep one has a probability-conserving transition matrix, which requires
\be 
\label{back}
\sum_{\Cp}
q(\Cp,\tau+\dt)
\langle \Cp| (\mident + \dt \WW(\g)) |\C\rangle q^{-1}(\C,\tau) = 1
\ee 
Given the reweighting factors $q(\Cp,\tau+\dt)$ at some time,  the factors $q(\C,\tau)$ at the previous timestep are therefore determined. If one collects the $q$-factors in a time-dependent vector $\langle q(\tau)| = \sum_\C \langle \C| q(\C,\tau)$, the relation~\eqref{back} reads
\be 
\langle q(\tau)| = \langle q(\tau+\dt) |
(\mident + \dt \WW(\g)) 
\ee 
or in the continuous time limit $\dt\to 0$
\be 
-\partial_\tau \langle q(\tau)| = \langle q(\tau)|\WW(\g) \, ,
\label{q_DE}
\ee
which is the \emph{backward} master equation for the biased master operator. Together with the final-time boundary condition $\langle q(t)|=\proj$ one gets the formal solution
\be 
\langle q(\tau)| = \langle e| \eee^{\WW(\g)(t-\tau)} \, .
\label{q_general_solution}
\ee
Once $\langle q(\tau)|$ has been found, the auxiliary master operator $\WW\aux(\tau)$ can be determined. For this, we use that the contribution from each timestep in  (\ref{equ:path_prob_multiplicative}) must equal $\langle \C_{\tau+\dt}| (\mident+\dt\WW\aux(\tau))|\C_\tau\rangle$.  Dividing both sides of the resulting equality by $\dt$, and simplifying the notation by replacing $\C_{\tau+\dt}\to \Cp$ and $\C_\tau \to \C$ we get
\begin{align} 
\frac{q(\C,\tau+\dt)-q(\C,\tau)}{q(\C,\tau)\dt}\delta_{\C,\Cp}+\frac{q(\Cp,\tau+\dt)}{q(\C,\tau)} \langle \Cp| \WW(\g) |\C\rangle = \langle \Cp| \WW\aux(\tau) |\C\rangle 
\end{align}
For $\dt\to 0$ this simplifies (using (\ref{q_DE})) to
\begin{align} 
\langle \Cp| \WW\aux(\tau) |\C\rangle &= 
\frac{q(\Cp,\tau)}{q(\C,\tau)} \langle \Cp| \WW(\g) |\C\rangle - \frac{\delta_{\C,\Cp}}{q(\C,\tau)} \sum_{\Cpp}q(\Cpp,\tau)\langle \Cpp|\WW(\g)|\C\rangle 
\label{equ:Waux}
\end{align}
Therefore the time-dependent auxiliary master operator $\WW\aux(\tau)$ is obtained by reweighting all transition rates of the unbiased master operator by an additional multiplicative correction, i.e.
\be 
% W\aux(\C\to\Cp, \tau) = W(\C\to\Cp) \eee^{-\g\alpha(\C,\Cp)} \frac{q(\Cp,\tau)}{q(\C,\tau)}
W\aux(\C\to\Cp, \tau) = W(\C\to\Cp) \eee^{-\g\alpha(\C,\C')} \frac{q(\Cp,\tau)}{q(\C,\tau)} 
\label{aux_transition_rates}
\ee
for all $\C \neq \Cp$. Equation~\eqref{aux_transition_rates} corresponds to equation~(9) in the manuscript. 

Before proceeding we comment briefly on our notation for the ``backwards factor'' $q$. From equation~\eqref{q_general_solution}, it is clear that the dependence of $q$ on $\tau$ is via the time difference $t - \tau$. We generally leave the $t$-dependence implicit and write %However, for simplicity, we would mainly write 
either $\langle q(\tau)|$ or $q(\C, \tau)$ as used above. For concrete calculations below in the thermodynamic limit, where we will need to perform Laplace transforms with respect to the variable $t - \tau$, we will use the notation $q(E, \tau) = Q(E, t- \tau) $ where appropriate.

Having established the auxiliary transition rates in~\eqref{aux_transition_rates}, it now remains to find the auxiliary escape rates, i.e.\ the (negative) diagonal terms of the auxiliary master operator. These follow by combining equations~\eqref{q_DE} and~\eqref{equ:Waux} as 
 %After obtaining the auxiliary transition rates~\eqref{aux_transition_rates} we go for the auxiliary escape rates. These are obtained in the usual way by summing the auxiliary transition rates, as the second term on the right hand side of (\ref{equ:Waux}) shows. An alternative form of them is
\be 
r_\tau\aux(\C) = -\langle \C|\WW\aux(\tau)|\C\rangle = 
-\langle \C|\WW(\g)|\C\rangle - \partial_\tau \ln q(\C,\tau)
\label{exit_rates}
\ee
%which can be seen using again equation~(\ref{q_DE}). 

With the above results in hand, we proceed to show the relation between the mean auxiliary escape rate and the time-dependent activity (equation~(12)). The argument is in fact the same as for standard Markov jump processes, but we include it here for completeness. Let us write the mean contribution to $\mathcal{A}$ from jumps between time $\tau=k\dt$ und $\tau+\dt$ as
% (associated with the time step $k$) as
\begin{align}
  \langle \alpha(\C_k, \C_{k+ 1}) \rangle &=   \sum_{\C_{k+1},\C_k}  P_s(\C_k, \C_{k+1})  \alpha(\C_k, \C_{k+1}) 
  \label{jumps}
\end{align}
Here the weighting is with the joint probability  $P_s(\C_k, \C_{k+1})$ in the biased process of the two configurations at the beginning and end of the time interval, $\C_k$ and $\C_{k+1}$. This probability is $P_s(\C_k, \C_{k+1}) = P_s(\C_k,\tau)\,\dt\, W^{\rm{aux}}(\C_k \to \C_{k+1}, \tau)$ (provided the two configurations are distinct; if they are identical then there is no contribution to~\eqref{jumps} anyway). It is for this step that the probability-conserving nature of $\WW^{\rm aux}(\tau)$ is crucial; had we used $\WW(s)$, the sum over the later states in the trajectory would have given an additional nontrivial factor, which in fact is proportional to $q(\C_k,\tau)$.

Inserting the expression for $P_s(\C_k, \C_{k+1})$, the mean contribution to our trajectory observable $\mathcal{A}$  between timesteps $k$ and $k+1$ is
\begin{align}
\langle  \alpha(\C_k, \C_{k+1}) \rangle
  &= \sum_{\C_{k+1},\C_k} P_s(\C_k,\tau)  W^{\rm{aux}}(\C_{k} \to \C_{k+1}, \tau)   \alpha(\C_k, \C_{k+1}) \, dt
%  &= \sum_{k =0} \sum_{\C_k} P_s(\C_k, \tau) r^{\rm{aux}}_\tau(\C_k) dt 
\end{align}
For the activity, where $\alpha(\C,\C')=1-\delta_{\C,\C'}$, the sum over $\C_{k+1}$ just gives the escape rate. Summing also over all timesteps up to $K=t/\dt$ to get the total activity $\mathcal{A}$ one has
% up to $K$ timesteps (corresponding to time $t=K\dt$) is
\begin{align}
  \langle \mathcal{A} \rangle_s &= 
%\sum_{k=0}^K \langle  \alpha(\C_k, 
%\C_{k+1}) \rangle \\
%  &= \sum_{\C_{k+1},\C_k} P_s(\C_k,\tau)  W^{\rm{aux}}(\C_{k} \to \C_{k+1}, \tau)  \notag \\
%  &\times \alpha(\C_k, \C_{k+1}) \, dt \\
%&= 
\sum_{k =0}^{K-1} \sum_{\C_k} P_s(\C_k, \tau) r^{\rm{aux}}_\tau(\C_k) dt 
\end{align}
In the continuous time limit this reads
\begin{align}
  \langle \mathcal{A} \rangle_s = \int_0^t d\tau\, \langle r^{\rm{aux}}_\tau \rangle_s  \equiv  \int_0^t d\tau\,\langle a_\tau \rangle_s
  \label{acti}
\end{align}
where the last equivalence is just the definition of the time-dependent activity $a_\tau$. Overall, one sees that the mean time-dependent activity $\langle a_\tau\rangle_s$ is just 
%The last equivalence comes from equation~\eqref{activ}. Thus,  
$\langle r^{\rm{aux}}_\tau \rangle_s$, the average escape rate (calculated from the auxiliary master operator) in the biased process. This demonstrates equation~(12) in the manuscript.

%An interpretation of the $q$-factors can be obtained by considering the probability in the biased ensemble of being in configuration $\C$ at time $\tau$.

For concrete calculations based on the above formalism we need an expression for the probability (again in the driven process) of being in configuration $\C$ at time $\tau$, $P_s(\C, \tau)$. If we sum the trajectory probability~\eqref{equ:path_prob_multiplicative} over the configurations at all other times we get 
\begin{align} 
  P_s(\C,\tau) &= \frac{ \proj \eee^{\WW(\g)(t-\tau)} |\C\rangle \langle \C| \eee^{\WW(\g)\tau} | P(0)\rangle}{Z(\g,t)}
%=  \frac{q(\C, \tau) \Pf(\C,\tau)}{Z(\g,t)}
\label{P_C_tau_dependent}
\end{align}
The first factor in the numerator is $q(\C,\tau)$; we define the second one as $\Pf(\C,\tau)$. To obtain eq.~(13) in the main text we just need to show that this solves the forward master equation for $\WW(\g)$. But this follows directly because 
\be
\Pf(\C,\tau)=\langle\C|\Pf(\tau)\ \ \mbox{with}\ \   
|\Pf(\tau)\rangle = \eee^{\WW(\g)\tau}| P(0)\rangle
\label{forward_probability}
\ee
and because
\be
% \Pf(\C,\tau) = \langle \C| \eee^{\WW(\g)\tau}| P(0)\rangle
{\partial_\tau} | \Pf(\tau) \rangle = \WW(\g)  | \Pf(\tau) \rangle
\label{forward_me}
\ee
Concerning the physical interpretation, one sees that $\Pf(\C,\tau)$ contains the effect of the bias propagated forwards from the past. Note that even though $\Pf(\C,\tau)$ is calculated from $\WW(\g)$ and the initial distribution $P(\C,0)$ exactly in the same way as for conventional Markov dynamics, it is not a normalized probability distribution because $\WW(\g)$ does not conserve probability.

The factor $q(\C, \tau)$ can now be interpreted similarly as the effect of the bias acting in the future, propagated backwards to the current time. Equation~\eqref{P_C_tau_dependent} shows that the actual probability of being in configuration $\C$ is simply a product of these past and future factors, suitably normalized by $Z(\g,t)$.  From equations~\eqref{equ:zmat} and~\eqref{P_C_tau_dependent} one also deduces that the normalization factor is nothing but
\begin{align}
  Z(s,t)  &= \sum_{\C} \Pf(\C, t)
            \label{partif_c}
\end{align}
%cf.~eq.\eqref{parte}. 
%Interpretation of $q$
%A useful way to think of the $q$-factors is in terms of an effective potential, defined by 
%\be 
%V\eff(\C,\tau) = -(2/\beta) \ln q(\C,\tau)
%\ee
%The transition rates (\ref{aux_transition_rates}) in the auxiliary master operator then take the Metropolis-like form
%\be 
%W\aux(\C\to\Cp) = W(\C\to\Cp) \eee^{-\g\alpha(\C,\Cp)} \eee^{-\beta [V\eff(\Cp,\tau)-V\eff(\C,\tau)]/2}
%\ee
%so that $V\eff(\C,\tau)$ can be identified as an effective additional contribution to the energy of each configuration $\C$. 

 %Explain how the reweighting becomes time-independent if no aging

Having set out the general construction for driven aging systems, we describe briefly how  %the auxiliary master operator 
it simplifies in the conventional scenario where the driven (biased) dynamics reaches a steady state because the biased master operator has a gap in its spectrum below the leading eigenvalue $\psi(s)$. In such a scenario, the exponential of $\WW(\g)$ is dominated by the largest eigenvalue in~\eqref{spectral} at times longer than the inverse spectral gap. This largest eigenvalue is $\psi(s)$, so that
\begin{align} 
Z(\g,t) &= \proj V_0\rangle \eee^{\psi(\g) t} \langle U_0|P(0)\rangle \label{partisteady} , \\
  \langle q(\tau)| &= \proj V_0\rangle \eee^{\psi(\g)(t-\tau)} \langle U_0| ,
  \label{qtau} \\
|\Pf(\tau)\rangle &= | V_0\rangle  \eee^{\psi(\g)\tau} \langle U_0|P(0)\rangle
\label{equ:TTI_results}
\end{align}
with $U_0$ and $V_0$ the top left and right eigenvectors of the biased master operator $\WW(s)$. Bearing in mind the eigenvector normalization $\langle U_0|V_0\rangle = 1$, one thus retrieves the known~\cite{garrahan2009first, jack2010large}  $\tau$-independent steady state distribution $P_s(\C,\tau) = \langle U_0|\C\rangle \langle \C|V_0\rangle$. One also finds that the escape rate shift $-\partial_\tau \ln q(\C,\tau)= \psi(\g)$ is just the dynamical free energy,
\begin{align}
  r\aux(\C) = r(\C) + \psi(s)
  \label{rmle_eq}
\end{align}
which is in agreement with previous results~\cite{evans2004detailed, evans2004rules} (see also~\cite{jack2010large}) and leads to~(14) in the main text.

Summarizing, in this section we have extended existing results for the representation of driven $s$--ensemble dynamics in terms of a physical (``auxiliary'') master operator
to the case where the $\tau$-dependence of the latter cannot be ignored, for instance because of the existence of aging effects. 
The key identities are~\eqref{aux_transition_rates} and~\eqref{exit_rates} for the elements of the auxiliary master operator, the equations~\eqref{q_DE} and~\eqref{forward_me} governing the backward and forward factors, the resulting expression for the $\tau$-dependent configuration probabilities~\eqref{P_C_tau_dependent} and finally, for the important case of driving by activity bias, the expression~\eqref{acti} for the total and time-dependent activities.

We note finally that the formalism applies without changes also if one adds to the biasing observable a time-integrated contribution of the form
\be 
\dt \sum_k b(\C_k) = \int_0^t d\tau\,b(\C_\tau)
\ee
The biased master operator then merely acquires an additional contribution of $-s\,b(\C)$ to its diagonal element $\langle \C|\WW(\g)|\C\rangle$, with a corresponding additional term $+s\,b(\C)$ on the r.h.s.\ of%the appropriate minus sign also appears in
~\eqref{rmle_eq}.

\subsection{Thermodynamic limit and top eigenvector systems}

In this subsection we show how to rewrite the master equation~\eqref{master} in the thermodynamic limit of an infinite number of traps identified by their depth $E$, which in the limit becomes a continuous variable; the result is eq.~(3) in the main text. We also derive the thermodynamic limit version of the forward (eq.~\eqref{forward_me}) and backward (eq.~\eqref{q_DE}) master equations. We additionally write the discrete and continuous version of the top eigenvalue system for the left and right eigenvectors.

As in the main text we assume that each configuration (or trap) $\C$ has an energy (trap depth) $E_\C$. The transition rates are taken to be functions of the energies only and we normalize them so that the escape rates in a system of $N$ traps with mean-field connectivity are $O(1)$, hence
\begin{align}
  W(\Cp \to \C) = \frac{1}{N} W(E_\Cp \to E_\C)
  \label{trans_energ}
\end{align}
where the difference between the Bouchand and BM models lies only in the function $W(E\to E')$.
We introduce the probability for the system to be in a trap with depth $E$ at time $\tau$ as $P(E,\tau) = \sum_\C P(\C, \tau) \delta(E - E_\C)$ and rewrite equation~\eqref{master} as an equation for $P(E,t)$. This only requires multiplying both sides by $ \delta(E - E_\C)$ and summing over all configurations:
  \begin{align}
%     \frac{\partial}
{\partial_\tau} P(E, \tau) &= -\frac{1}{N} \sum_{\C, \Cp \neq \C }  W(E_\C \to E_\Cp) P(\C,\tau) \delta(E - E_\C)  + \frac{1}{N} \sum_{\C, \Cp \neq \C } W(E_\Cp\to E_\C) P(\Cp,\tau) \delta(E - E_\C)
\end{align}
The constraint on the sums can safely removed without changing the equality as the terms with $\C=\Cp$ cancel. In the transition rate factor of the last term we can replace $E_\C$ by $E$ because of the Dirac delta; the remaining sum over $\C$ then gives the density of states $\rho(E) = \frac{1}{N} \sum_\C \delta(E - E_\C)$:
%, we also point out that we  can change one of the arguments of the transition rates  $W$ according to the value prescribed by the Dirac's delta and then perfom the sum over $\C$:
\begin{align}
       \frac{\partial}{\partial \tau} P(E, \tau)   &= -\frac{1}{N}  \sum_{\Cp }  W(E \to E_\Cp) P(E, \tau)  + \sum_{\Cp} W(E_\Cp\to E) P(\Cp,\tau) \rho(E)
    \label{finitemaster}
\end{align}
%In the second term we have used the density of states, defined as . In the thermodynamic limit, $N \to \infty$, we replace $\rho_N(E) \to \rho(E)$. 
Finally, we introduce a factor of $1 = \int dE'\, \delta(E' - E_\Cp)$  on the r.h.s.\ of equation~\eqref{finitemaster} and perform the sum over $\Cp$. The result is
\begin{align}
%    \frac{\partial}
    {\partial_\tau} P(E, \tau) &= - r(E) P(E, \tau)  + \rho(E) \int dE'  P(E',\tau) W(E' \to E) 
\end{align}
which corresponds to equation~(3) in the main text, 
with the escape rate
\be
r(E) = \frac{1}{N} \sum_{\Cp} W(E\to E_{\Cp}) = \int dE'\,W(E\to E')\rho(E') 
\ee
Note that the escape rate for finite $N$ is $r_\C = 1/N \sum_{\Cp\neq \C} W(E_\C\to E_{\Cp})$. Because of the restriction $\Cp\neq \C$ this differs from $r(E)$ by a term of $O(1/N)$, but this becomes irrelevant in the thermodynamic limit $N\to\infty$. Such corrections are ignored by default in all derivations that follow.

The calculation for the continuous limit of the forward master equation (eq.~\eqref{forward_me}) is essentially the same as for the unbiased case. One only has to consider the biased master operator as defined in~\eqref{equ:WA} instead of the conventional one. The result is, for $N\to\infty$,
\begin{align}
%    \frac{\partial}
    {\partial_\tau} \Pf(E, \tau) &= - r(E) \Pf(E, \tau)  + {\rm{e}}^{-s} \rho(E) \int dE'  \Pf(E',\tau) W(E' \to E)
                                                  \label{pf_continuum}
\end{align}
For the backward equation (eq.~\eqref{q_DE}) we show the calculation in more detail because some care is required in the definition of the thermodynamic limit of the backward factor:
% we now do the explicit calculation.  Before starting, let us point out the way in which the factor $q$ transforms, namely:
\begin{align}
  q(E, \tau) \rho(E) = \frac{1}{N} \sum_\C q(\C, \tau) \delta (E - E_\C)
  \label{q_transform}
\end{align}
The factor $\rho(E)$ on the l.h.s.\ is needed because in the unbiased case the backward factor should reduce to a unit constant, $q(E,\tau)=q(\C,\tau)=1$, corresponding to the conservation of probability. From~\eqref{q_general_solution} the same has to hold generally, even in the presence of a trajectory bias, at the final time $\tau=t$.
%This means that we consider $q$ as a scalar function. A way to see the validity of equation~\eqref{q_transform} is by evaluating the previous expression at the boundary $\tau = t$. From the formal solution~\eqref{q_general_solution}, we know that $q(\C, t) = 1$ and hence $q(E, t) = 1$, therefore the right hand side of eq.~\eqref{q_transform} would be nothing but $\rho(E)$. 
We now start from equation~\eqref{q_DE}, insert the decomposition $\langle q(\tau) | = \sum_\C q(\C, \tau) \langle C |$
%:
%\begin{align}
%  -    \sum_\C   %\frac{\partial}
%  {\partial_\tau} q(\C, \tau)  \langle \C | = \sum_\C q(\C, \tau) \langle \C | \WW(s) 
%\end{align}
and multiply by $| \Cp \rangle$ from the right to project onto configuration $\Cp$:
%And let us project this expression into the state 
\begin{align}
  -    \sum_\C  %\frac{\partial}
  {\partial_\tau} q(\C, \tau)&  \langle \C | \Cp \rangle =  \sum_\C q(\C, \tau)  \langle \C | \WW(s) | \Cp \rangle \\
  -
  %\frac{\partial}
  {\partial_\tau} q(\Cp, \tau) &=  \sum_\C q(\C, \tau) \big[ W(\Cp \to \C) {\rm{e}}^{-s} (1 - \delta_{\C, \Cp}) 
  %\notag \\  &
  - r(\Cp)\delta_{\C, \Cp}  \big] \\
&=  \sum_{\C,\C\neq \Cp} q(\C, \tau) W(\Cp \to \C) {\rm{e}}^{-s} %(1 - \delta_{\C, \Cp})
  - q(\Cp,\tau) \sum_{\C,\C\neq \Cp} W(\Cp\to \C)
\end{align}
where in the last line we have used the definition (for finite $N$) of the escape rate $r(\Cp)$.
Now express %Then we replace the expressions for 
the transition rates in terms of energies (eq.~\eqref{trans_energ}), multiply both sides of the equation by $\delta(E - E_\Cp)/N$ and sum over all configurations $\Cp$. This yields
\begin{align}
  &-  \rho(E)    {\partial_\tau} q(E, \tau) = \frac{ {\rm{e}}^{-s} }{N^2}    \sum_{\C \neq \Cp} q(\C, \tau) W(E_\Cp \to E_\C) \delta(E - E_\Cp) -  \frac{1}{N^2} \sum_{\C \neq \Cp} q(\Cp, \tau) W(E_\Cp \to E_\C) \delta(E - E_\Cp)
\end{align}
The constraint $\C \neq \Cp$ only gives $O(1/N)$ corrections so can be dropped; % as the contribution from $\C = \Cp$ is immaterial for $N \to \infty$  
performing the sum over $\Cp$ then yields
\begin{align}
  -  \rho(E)   
  %\frac{\partial}
  {\partial_\tau} q(E, \tau) =   {\rm{e}}^{-s}   \frac{\rho(E)}{N}    \sum_{\C } q(\C, \tau) W(E \to E_\C) - \frac{1}{N} q(E, \tau) \rho(E) \sum_\C W(E \to E_\C)
\end{align}
Finally we cancel the common factor of $\rho(E)$ and again switch to an integration over trap depths on the r.h.s.\ by inserting $1 = \int dE'  \delta(E' - E_\C)$ and performing the remaining sum over $\C$:
%\begin{align}
 % -   \frac{\partial}{\partial \tau} q(E, \tau) &=  \frac{{\rm{e}}^{-s}}{N} \int dE' W(E \to E') \sum_\C q(\C, \tau) \delta(E' - E_c) \notag \\ &- q(E, \tau) \frac{1}{N} \int dE' W(E \to E') \sum_\C  \delta(E' - E_c) 
%\end{align}
%The result after  is:
\begin{align}
-   % \frac{\partial}
{\partial_\tau} q(E, \tau) &= 
{\rm{e}}^{-s}  \int dE'  q(E',\tau) W(E \to E') \rho(E') - q(E, \tau) \int dE'\, W(E\to E') \rho(E') 
                                                 \label{final_qe}
\end{align}
As the final integral is just $r(E)$, this is equation~{(10)} in the main text.

In the explicit calculations below we will often work with Laplace transforms. As $q(E, \tau)$ depends on the trajectory length $t$ and the running time $\tau$ only via the time difference $\Delta t = t - \tau$ (cf.\ eq.~\eqref{q_general_solution}), it will then be helpful to work with the quantity $Q(E, \Delta t) = q(E, \tau)$, for which equation~\eqref{final_qe} becomes
\begin{align}
  \frac{\partial}{\partial \Delta t} Q(E, \Delta t) &= - r(E) Q(E, \Delta t) + {\rm{e}}^{-s}  \int dE'  Q(E',\Delta t) W(E \to E') \rho(E')
                                                      \label{final_Qe}
\end{align}

We finally write down, for later reference, the equations for the top left and top right eigenvectors of the biased master operator, in the discrete case (finite $N$) and in the continuous ($N\to\infty$) limit. These eigenvectors correspond directly to solutions of the backward and forward master equations, respectively, 
%with an exponential time dependence $\propto {\rm{e}}^{\psi(s)\tau}$ 
with the time derivatives replaced by the top eigenvalue, which according to eq.~\eqref{equ:res-psi} is the dynamical free energy $\psi(s)$.
% multiplied by the corresponding eigenvector. 
For the top right eigenvector we use the notation $V_0(\C)$ in the discrete case. The eigenvalue problem then reads
\begin{align}
   &- V_0 (\C) r(\C) +  {\rm{e}}^{-s} \sum_{\Cp \neq \C} V_0(\Cp) W(\Cp \to \C)  = \psi(s)  V_0(\C) 
\end{align}
and becomes in the continuous limit (using that $V_0(E) = \sum_\C \delta(E - E_\C) V_0(\C) $)
%, one gets
\begin{align}
      -V_0(E) r(E) &+ {\rm{e}}^{-s} \rho(E) \int dE' W(E' \to E) V_0(E')                    = \psi(s)  V_0(E)
                     \label{rightE}
\end{align}
Notice that the left hand side is exactly the same as the right hand side of equation~\eqref{pf_continuum} with $\Pf$ replaced by $V_0$, as it should be.

For the left eigenvector $U_0(\C)$ we have the following condition for a discrete set of states
\begin{align}
    - U_0 (\Cp) r(\Cp) + \sum_{\C \neq \Cp} U_0(\C) W(\Cp \to \C) {\rm{e}}^{-s}  = \psi(s)  U_0(\Cp) 
\end{align}
The continuous limit reads (with $U_0(E) \rho(E) = 1/N \sum_\C \delta(E - E_\C) U_0(\C) $)
\begin{align}
      -U_0(E) r(E) &+ {\rm{e}^{-s}} \int dE' \rho(E')  W(E \to E') U_0(E') = \psi(s)  U_0(E)
                     \label{contin_lefte}
\end{align}
with the l.h.s.\ again matching\ eq.~\eqref{final_qe}. Note that because the activity is invariant under time-reversal and because the undriven dynamics obeys detailed balance, i.e.\ $e^{\beta E}W(E\to E') = e^{\beta E'} W(E'\to E)$, the left and right eigenvector problems can be related to each other and one finds in particular that $V_0(E) \propto U_0(E)\rho(E)e^{\beta E}$.

\subsection{Bouchaud trap model}

% Next paragraph may be added to the main text
In this section we provide calculational details for the activity--biased dynamics in the Bouchaud trap model. We start for completeness with the calculation of the mean time--averaged activity in the unbiased case, which will act as a baseline for the results in the presence of driving by trajectory bias.

We note that as the Bouchaud trap model is a renewal process, some of our results can be related to those from the classical reference~\cite{godreche2001statistics}. We will comment on the specific instances of this connection below. The link 
%between the parametrizations of the problem
is made via the distribution of interarrival times, i.e.\ times between jumps, which is given by
\begin{align}
    \rho(\tau) = \int_1^\infty d\tilde{\tau} \,\frac{1}{\tilde{\tau}}\exp(-\tau/\tilde{\tau}) \tilde{\rho}(\tilde{\tau}) 
    \label{intera}
\end{align}
in terms of $ \tilde{\rho}(\tilde{\tau}) = T \tilde{\tau}^{-(T+1)}$, the mean trapping time distribution (which in turn follows from $\rho(E)=\eee^{-E}$ and $\tilde\tau = 1/r(E)=\eee^{E/T}$). Evaluating the integral in~\eqref{intera} gives
\begin{align}
 \rho(\tau) = T \tau^{-(T+1)} [T \Gamma(T) - \Gamma(1+T, \tau)]
\end{align}
with $\Gamma(\cdot)$ and $\Gamma(\cdot, \cdot)$ the complete and (upper) incomplete Gamma function, respectively. The latter can be neglected asymptotically and a comparison with equation~(1.6) from reference~\cite{godreche2001statistics} then reveals that the mapping between the notation there and here is $\theta \mapsto T$, $\tau_0 \mapsto (T \Gamma(T))^{1/T}$.
%We will point out the specific connections of our results to those of ~\cite{godreche2001statistics}

%We have checked in particular the result for the mean time--averaged activity (equation~\eqref{zeroscaling}) and the dynamical free energy around $s = 0$ (see equations~\eqref{free1} and~\eqref{free2} below). %One further comment is in order, though. From reference~\cite{godreche2001statistics} we use the asymptotic distribution for the number of renewals (for large times) as starting point to compute $\psi(s)$ and then the steepest descent method to evaluate the mean value of $\exp(-s  \mathcal{A}_t)$. 

\subsubsection{Zero bias}

\begin{figure}
  \includegraphics[width=0.5\textwidth]{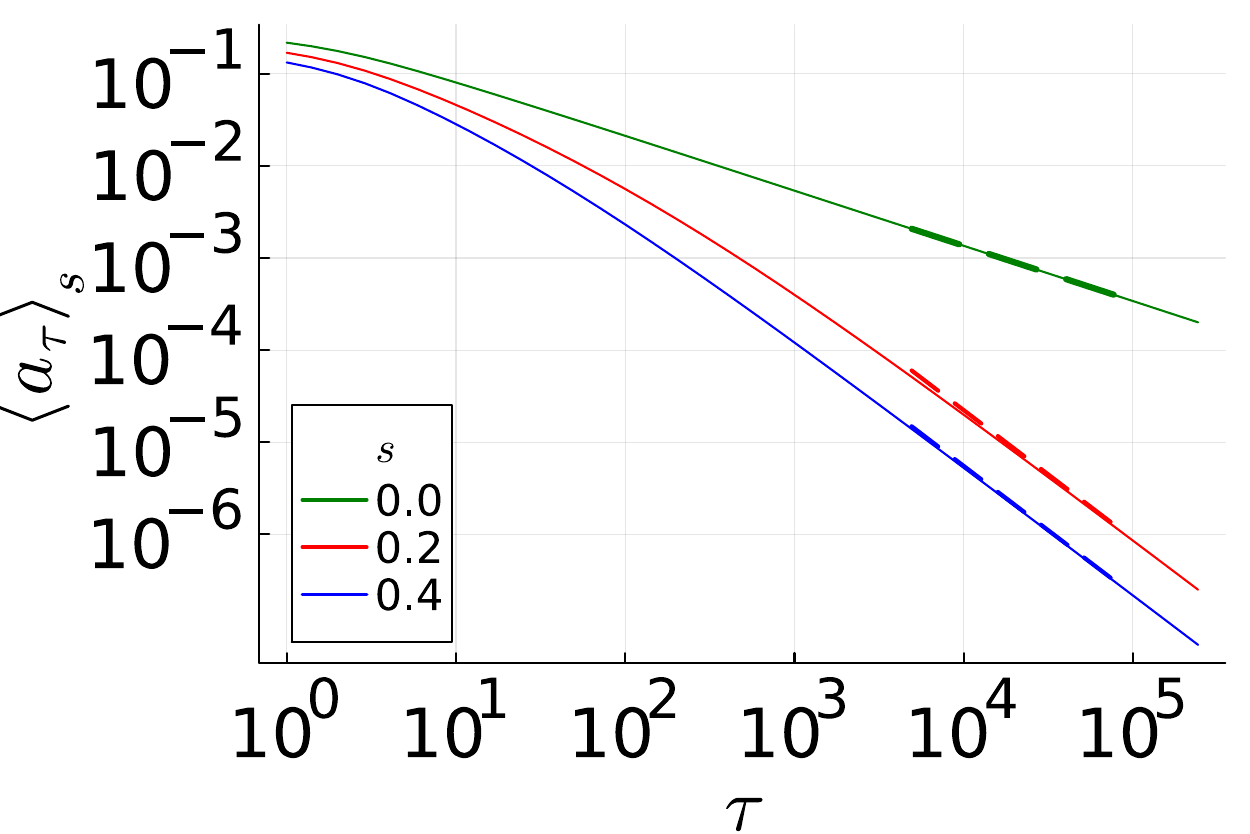}
  \caption{Evolution of the mean time--dependent activity with time for different biases at $T=0.4$ for the Bouchaud TM. Solid lines are from numerical solutions, {obtained as the time--domain version of equation~\eqref{pihat}}. The analytical solution for the unbiased case (eq.~\eqref{zeroscaling}) is shown by the green dashed line, while the red and blue dashed lines display the predictions~\eqref{pos_bias_scaling} for positive biases.  Total trajectory length (for the biased cases) $t = 10^6$. }
\label{activ_bou}
\end{figure}

Our starting point is the master equation~(3) with the appropriate transition rates (eq.~(1)) and escape rates $r(E)=W(E\to E')=\eee^{-E/T}$:
\begin{align}
%  \frac{\partial}
  {\partial_\tau} P(E,\tau) &= -  {\rm{e}}^{- E/T} P(E,\tau) + \rho(E) \int_0^\infty dE'\, {\rm{e}}^{-E'/T} P(E',\tau) 
\end{align}
We consider throughout a uniform distribution across traps as initial condition, corresponding to $P(E,0) = \rho(E) = {\rm{e}}^{-E}$. Solving the dynamical equation in the Laplace domain~\cite{MonBou96}, with $z$ conjugate to $\tau$, a straightforward calculation yields
\begin{align}
   \hat{P}(E,z) &= \frac{{\rm{e}}^{-E}}{(z+{\rm{e}}^{-E/T})f(z)}
   \label{Phat_unbiased}
\end{align}
with (all integrations over $E$ run from 0 to $\infty$ in what follows)
\be 
f(z) = \int%_0^\infty 
dE\, \frac{z\,{\rm{e}}^{-E}}{z+{\rm{e}}^{-E/T}} = {}_2F_1(1, T, 1+T, -1/z)
\label{f_def}
\ee
which as shown can be expressed in terms of the Gauss hypergeometric function $%F\equiv 
{}_2F_1$. The mean time--dependent activity, which is equal to the mean escape rate by eq.~\eqref{acti}, is 
\be 
\langle a_\tau\rangle_0 = \int dE\,P(E,\tau)\eee^{-E/T}
\ee
and has Laplace transform
\be 
\langle \hat a(z)\rangle_0 = \int dE\,\hat{P}(E,z)\eee^{-E/T} = \frac{1}{f(z)} \int dE\,\frac{{\rm{e}}^{-E-E/T}}{z+{\rm{e}}^{-E/T}} = \frac{1-f(z)}{f(z)}
\label{az_Bouchaud_unbiased}
\ee
% \frac{1}{z^T\, T \Gamma(T) \Gamma(1-T)}\int dE\,\frac{{\rm{e}}^{-E-E/T}}{z+{\rm{e}}^{-E/T}}
%\ee
While the above expressions are pleasingly explicit, the inverse Laplace transform cannot be obtained in closed form. We therefore focus on the physically most interesting limit of long times, $z \to 0$. With the variable change $x=\eee^{-E}$ we obtain
\be 
f(z) = \int_0^1 dx\,\frac{z}{z+x^\beta} 
\ee
where $\beta=1/T$ as usual. For temperatures below the glass transition, $T<1$, the substitution $y=x/z^T$ transforms this to
\be 
f(z) = z^{T} \int_0^{z^{-T}}\!\!\frac{dy}{1+y^\beta}
\ee
For $z\to0$ the upper integration boundary diverges and the integral evaluates to $\pi T/\sin(\pi T)$. Combining with the straightforward expansion for $T>1$, one has overall for small $z$ 
\be
f(z) \approx \left\{ 
\begin{array}{lcl}
z/(1-\beta) & \mbox{for} & T>1\\
z^{T} \frac{\pi T}{\sin(\pi T)} & \mbox{for} & T<1
\end{array}
\right.
\label{f_small_z}
\ee
Inserting into~\eqref{az_Bouchaud_unbiased} and transforming back to the time domain shows that for long times the activity $\langle a_\tau\rangle_0$ approaches the constant value $1-\beta$ when $T>1$, indicating a conventional equilibrium state. For $T<1$, on the other hand,
\iffalse
We simplify further the obtained expression by noticing that the escape rate $r(E) = {\rm{e}}^{-E/T}$ (cf. eq.~\eqref{ratebou}) appears naturally in the first term in the denominator. In the Laplace domain, deep traps are characterized by the condition $r(E) \ll z$, i.e. its characteristic rate is smaller than the measured rate, whereas the opposite condition $r(E) \gg z$ characterizes the shallow traps. From this and specially in the glassy phase, we can argue that the dominant contribution to the mean escape rate comes from the shallow traps and therefore we approximate eq.~\eqref{apbou} as
\begin{align}
    \hat{P}(E,z) \approx \frac{{\rm{e}}^{-E} {\rm{e}}^{E/T}}{z^T T \Gamma(T) \Gamma(1-T) }
\end{align}
The inverse Laplace transform of this function (with conjugate variable $\tau$ to be consistent with upcoming notation) is
\begin{align}
  P(E,\tau) \approx  \frac{\sin(\pi T)}{\pi T \Gamma(T)} {\rm{e}}^{-E} {\rm{e}}^{E/T} \tau^{T-1} 
\end{align}
And the mean escape rate (equal to the mean time--averaged activity, cf. eq.~\eqref{acti}) becomes
\fi
%
\begin{align}
  \langle a_\tau \rangle_0 
  %&= \int_0^\infty dE\,P(E,\tau) {\rm{e}}^{-E/T} \\
                           &\approx \frac{\sin(\pi T)}{\pi T \Gamma(T)}  \tau^{T-1}
                             \label{zeroscaling}
\end{align}
This prediction compares well against results from numerical solutions of the master equation as shown in Fig.~\ref{activ_bou}. The mean total activity can be obtained as the integral of the last expression (cf.\ eq.~\eqref{acti}), with the result
\begin{align}
  \langle \mathcal{A} \rangle_0 \approx  \frac{\sin(\pi T)}{\pi T^2 \Gamma(T)}  t^{T}
  \label{Bouchaud_A0}
\end{align}
which corresponds exactly to equation (3.6) in reference~\cite{godreche2001statistics}.

\subsubsection{Bias towards high activity}

Our starting point is the right eigenvalue equation~\eqref{rightE} with the appropriate transition and escape rates (see eq.~(1)) %and~\eqref{ratebou}). This is
\begin{align}
  \psi(\g) V_0(E) &= -{\rm{e}}^{-E/T} V_0(E) +  {\rm{e}}^{-s} \rho(E) \int dE' {\rm{e}}^{-E'/T} V_0(E') 
  \label{forwbou}
\end{align}
which may be rearranged to read
\begin{align}
  V_0(E) &=  \frac{{\rm{e}}^{-s} \rho(E) \int dE' {\rm{e}}^{-E'/T} V_0(E')}{\psi(\g) + {\rm{e}}^{-E/T}} 
%\end{align}
%and hence
%\begin{align}
%    V_0(E) &
\propto \frac{  \rho(E)}{\psi(\g)+ {\rm{e}}^{-E/T}}
           \label{rightev}
\end{align}
Inserting back into eq.~\eqref{forwbou} gives the condition
\be
\int dE\,\rho(E)\frac{\eee^{-E/T}}{\psi(\g)+\eee^{-E/T}} = \eee^{\g}
\label{meancond} 
\ee
For the case of an exponential density of states that we consider throughout, $\rho(E)=\eee^{-E}$ for $E>0$, the function on the l.h.s.\ becomes simply $1-f(\psi(s))$ with $f$ defined in~\eqref{f_def}, giving
\be
%\ftwo(\psi(s),\beta)=\eee^s \qquad \mbox{with} \qquad
%\ftwo(\lambda,\beta)=\int_0^1 dx\, \frac{x^\beta}{\lambda+x^\beta}
f(\psi(s))=1-\eee^{s}
\label{Bouchaud_eigenval_condition}
\ee
%and $\beta = 1/T$ as usual.
%, and we need to solve the equation .
The function $f(z)$ increases monotonically with $z$ from $f(0)=0$ to $f(z\to\infty)=1$; the condition~\eqref{Bouchaud_eigenval_condition} %\ftwo(\psi(s),\beta)=\eee^\g$ 
therefore has exactly one positive solution $\psi(\g)$ for all $\g<0$. The other eigenvalues are all negative as one can see from the finite-$N$ form of the eigenvalue condition given in~\eqref{finiteN} below. As the discussion there shows, the second eigenvalue approaches zero for $N\to\infty$ so that the spectrum has a gap, of size $\psi(s)$. 
%As the condition for all other eigenvalues has the same form -- with the dynamical free energy replaced by the corresponding eigenvalue -- they must all be negative, and there is therefore a gap in the spectrum of size $\psi(s)$. (If one wanted to find the negative eigenvalues explicitly one would need to use the  because $f(z)$ becomes undefined for negative real $z$.)

\begin{figure}
\includegraphics[width=0.8\textwidth]{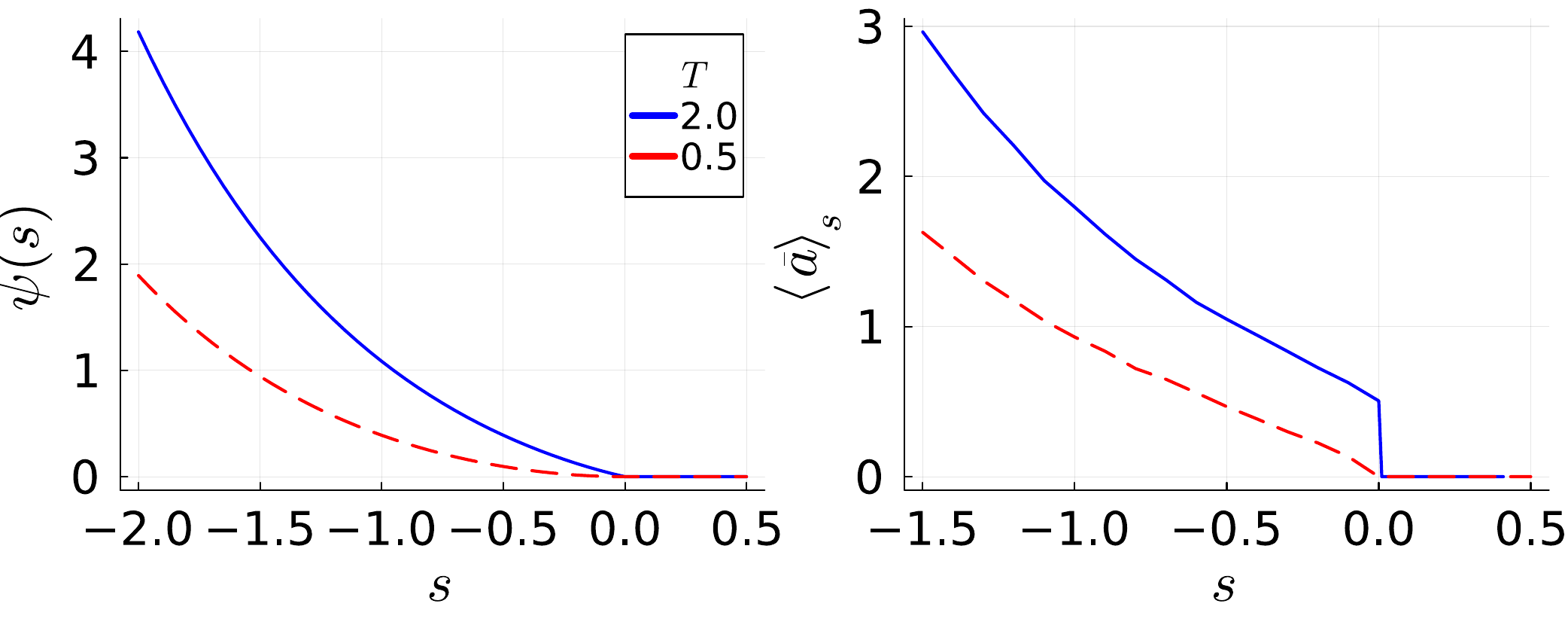}
\caption{(Left) Dynamical free energy in the Bouchaud trap model against bias parameter $\g$, for two different temperatures. (Right) Mean time--averaged activity. The kink at $\g=0$ shows a first order transition from an active ($\g<0$) to an inactive dynamical phase. {The free energy is obtained as the numerical solution of condition~\eqref{Bouchaud_eigenval_condition}. The activity curves are obtained by first--order perturbation theory as explained in the text.}} 
\label{fig:free_energy_Bouchaud}
\end{figure}

To summarize, for bias towards high activity ($\g<0$) the dynamical free energy of the Bouchaud model is obtained by solving equation~\eqref{meancond} with an exponential density of states, or explicitly~\eqref{Bouchaud_eigenval_condition}. The resulting free energy is plotted as a function of $\g$ for two different temperatures in Fig.~\ref{fig:free_energy_Bouchaud}. In the plot we have anticipated that the dynamical free energy goes to zero for $\g>0$, as we will show in the next subsection. The derivative $-\psi'(s)=\langle \bar{a}\rangle_s$ is nonzero for $\g<0$ so that we always obtain a nonzero time--averaged activity: stationarity in the Bouchaud model is restored by bias to high activity for all temperatures, even those ($T<1$) where the unbiased model is in its glassy phase and exhibits aging. 

Differences between the regime below and above the glass transition become visible, however, if we consider the behaviour of the dynamical free energy around $\g=0$. For small negative $\g$ we can expand the exponential in equation~\eqref{meancond} as ${\rm{e}}^{\g} = 1 - |\g|$, giving the condition $|s| = f(\psi(s))$. Using then the small-$z$ behaviour~\eqref{f_small_z} of $f(z)$ one finds 
\iffalse
 Inserting the ansatz $\psi(s) = b_0 |\g|^\alpha$ with a critical exponent $\alpha$ then %gives
% + o(|\g|^\alpha )$. 
transforms the condition~\eqref{Bouchaud_eigenval_condition} into
%We therefore write condition~\eqref{meancond} as
\begin{align}
     |\g|^{1-\alpha} =  b_0 \int_0^1 \frac{dx}{x^\beta + b_0 |\g|^\alpha} 
\end{align}
For $\beta < 1$, the r.h.s.\ has a finite limit for $|s|\to 0$ %term $x^\beta$ dominates the integral 
and it follows that $\alpha = 1$ and $b_0 = 1/\int_0^1 dx\,x^{-\beta} = 1-\beta $. For $\beta > 1$, on the other hand, the substitution $y = x(b_0 |\g|^{\alpha})^{-1/\beta}$ gives
\begin{align}
    |\g|^{1-\alpha} \approx  b_0^{1/\beta} |\g|^{\alpha/\beta - \alpha} \int_0^{(b_0 |\g|^\alpha)^{-1/\beta}} \frac{dy}{y^\beta + 1}
\end{align}
As $\alpha > 0$, the upper integration boundary goes to infinity in the $|\g| \to 0$ limit. The integral then becomes independent of $\g$ and one deduces that $\alpha = \beta$ and 
\begin{align}
    b_0 = \left( \int_0^\infty \frac{dy}{y^\beta + 1} \right)^{-\beta} = \left(\frac{\sin(\pi T)}{\pi T}\right)^{1/T}
    \label{a0}
\end{align}
Collecting the results we have 
\fi 
that around the phase boundary $\g = 0$, the dynamical free energy on the active side ($\g<0$) behaves as
 \begin{align}
  \psi(\g) &\approx (1 - \beta) |\g| \, , \qquad T> 1
  \label{free1}
  \\
  \psi(\g) &\approx b_0 |\g|^\beta \, , \qquad T<1
  \label{free2}
\end{align}
with $b_0=\left({\sin(\pi T)}/{\pi T}\right)^{1/T}$.
%given by equation~\eqref{a0}.
Equations~\eqref{free1} and~\eqref{free2} tell us that the dynamical phase transition is first order at high temperatures, but continuous in the temperature range where the unbiased system is glassy. Correspondingly, the time--averaged activity $\langle {a}\rangle_s = \psi'(\g)$ has a jump at the transition for $T>1$, while it approaches zero continuously for $T<1$ (cf. Fig.~\ref{fig:free_energy_Bouchaud}(right)).

Let us connect the above result with those of reference~\cite{godreche2001statistics}, focussing on the glassy regime $T<1$. One can exploit the G\"artner--Ellis theorem to compute the rate function $I(a)$ of the time--averaged activity $\bar{a}=\mathcal{A}/t$ in the long--time limit, see e.g.~\cite{jack2020ergodicity}, as
\begin{align}
  I(\bar{a}) = \sup_s [-s\bar{a} - \psi(s) ]
  % \sim a^{\frac{1}{1-T}}
\end{align}
Using now equation~\eqref{free2} (together with $\psi(s) = 0$ for $s > 0$, see next subsection), one finds directly that the scaling of the rate function around the minimum $a = 0$ must be (for $\bar{a}>0$)
\begin{align}
  I(\bar{a}) \sim \bar{a}^{\frac{1}{1-T}}
\end{align}
Thus to exponential order for large $t$, the distribution of the time--averaged activity (at zero bias) is
\begin{align}
  P(\bar{a}) \asymp \eee^{- t I(\bar{a})} \sim \eee^{-ct \bar{a}^{1/(1-T)}}
  \label{patotal}
\end{align}
with $\asymp$ indicating an equality up to subexponential prefactors and $c$ some $T$-dependent constant. In terms of the total activity, eq.~\eqref{patotal} translates into 
\begin{align}
  P(\mathcal{A}) \asymp \eee^{-c t^{-T/(1-T)} \mathcal{A}^{1/(1-T)}} = \eee^{-c (\mathcal{A}/t^{T})^{1/(1-T)}}
\end{align}
This scaling of the activity distribution corresponds exactly to that found for ``large'' activities in reference~\cite{godreche2001statistics} by a saddle-point approximation, see equation (3.8) and below in reference~\cite{godreche2001statistics}. Interestingly, the scaling regime considered there was that of activities of order of the average~\eqref{Bouchaud_A0}, i.e.\ $\mathcal{A}\sim t^T$, while in the large deviation rate function $I(\bar{a})$ one considers $\mathcal{A} \sim t$. What the comparison above shows, then, is that the result for the probability distribution of $\mathcal{A}$ in that regime (for large $\mathcal{A}/t^T$) agrees with that of our calculation for small $|s|$, i.e.\ small $\bar{a}=\mathcal{A}/t$. In other words, the two approaches match for $t^T \ll \mathcal{A} \ll t$ where the relevant scaling regimes meet. Note that our numerical results for the dynamical free energy for large $|\g|$, where the asymptotic expression~\eqref{free2} no longer applies, have no analogue in reference~\cite{godreche2001statistics} as they relate to larger activity values with $a=\mathcal{A}/t$ of order one.

%It is remarkable, nevertheless, that our approach based on the large deviation function considers explictly that $A$ is an extensive variable (in $t$), whereas the approach of reference~\cite{godreche2001statistics} considers a limiting distribution for the variable $A/t^T$. Both approaches naturally coincide as long as  $A/t^T \gg 1$. 
\begin{figure}
  \includegraphics[width=0.8\textwidth]{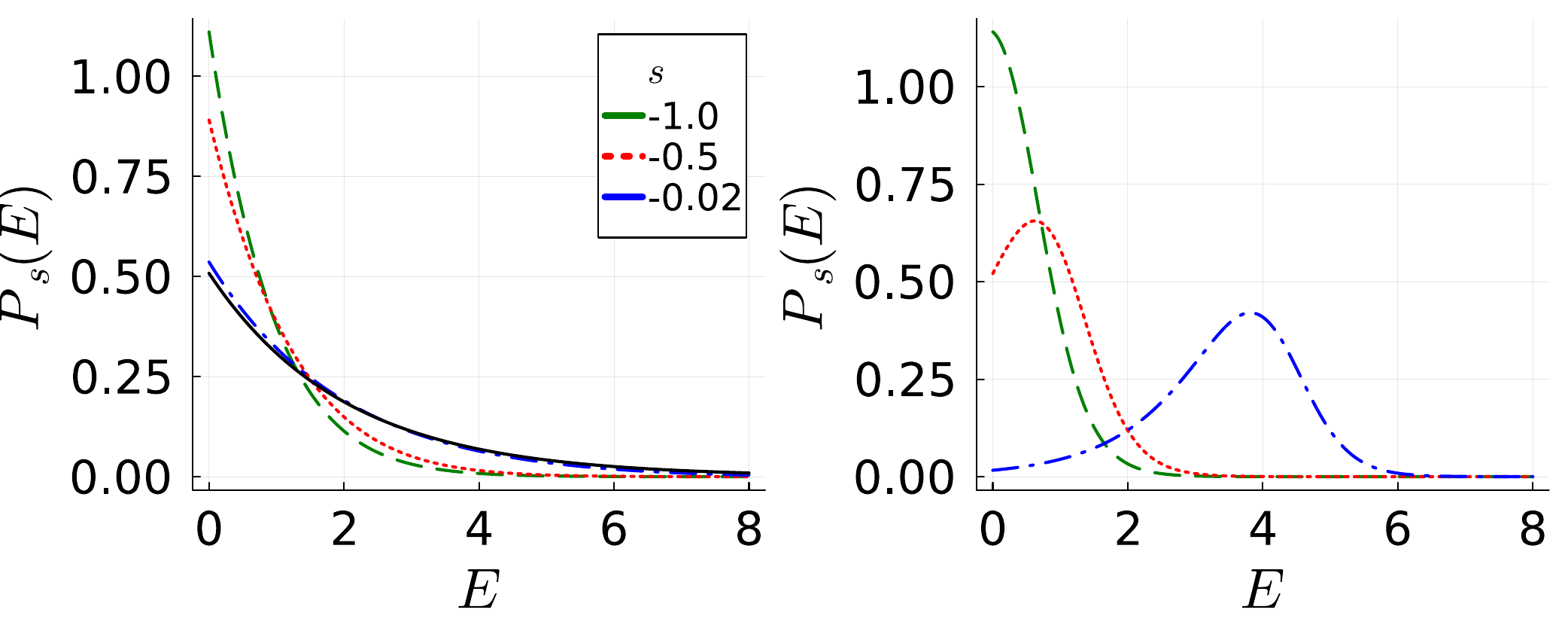}
  \caption{Steady-state trap depth distributions (eq.~\eqref{P_E_active_Bouchaud}) in the biased Bouchaud trap model for different values of $s$. Left: $T=2$, with corresponding equilibrium distribution shown as a solid line. Right: $T=1/2$, showing that the distribution develops a maximum around increasingly deep traps as $|\g|$ decreases.
\label{fig:Pss_active_Bouchaud}
}
\end{figure}

We can gain further insight into the behaviour for bias towards high activity by studying the steady state distribution of trap depths. As explained after (\ref{equ:TTI_results}), the steady state probability for a given configuration is $P_s(\C) = \langle U_0|\C\rangle \langle \C|V_0\rangle$, and correspondingly in the thermodynamic limit $P_s(E) = U_0(E) V_0(E)$.  
We have already determined the right eigenvector $V_0(E)$; the left eigenvector $U_0(E)$ can be found similarly starting from equation~\eqref{contin_lefte} or, as discussed below that equation, one can use directly that $U_0(E)\propto V_0(E)/[\rho(E)e^{\beta E}]$. % similar calculation to the one for the right eigenvector one can compute the top left eigenvector.
Either way one finds for the steady state distribution
\be
P_s(E) \propto \frac{\eee^{-(1-\beta)E}}{(1+\psi(s) \eee^{\beta E})^2}
\label{P_E_active_Bouchaud}
\ee
and some example plots are shown in Fig.~\ref{fig:Pss_active_Bouchaud}. While negative bias leads to active phases for all $T$ as we saw above, the behaviour of $P_s(E)$ as $\g\to0$ shows the signature of the glass transition: for $T>1$, $P_s(E)$ smoothly approaches the Boltzmann equilibrium distribution of the unbiased system; in the glass regime $T<1$, on the other hand, $P_s(E)$ develops a maximum that shifts to arbitrarily large trap depths as $\g\to 0$. This makes sense physically: without bias, the system ages into deeper traps over time. Any high activity bias stops this aging, but it does so in increasingly deep regions of the landscape as the bias gets weaker.

\subsubsection{Bias towards low activity}

We next look at the effect of biasing towards lower values than typical activity using a bias parameter $\g>0$. That the dynamical free energy in this regime tends to zero for $N\to\infty$ can be seen easily starting from the analogue of the condition~\eqref{meancond} for a finite number $N$ of traps, which reads 
\begin{align}
  \frac{1}{N} \sum_{\C} \frac{\eee^{-E_\C/T}}{\psi(\g)+\eee^{-E_\C/T}} = \eee^{\g}
  \label{finiteN}
\end{align}
 Let us assume that the trap depths are arranged in increasing order: $0<E_1<\ldots<E_N$. It is clear that the left hand side of equation~\eqref{finiteN} as a function of $\psi$ has poles at $\{- \eee^{-E_\C/T} \}$, and therefore the largest eigenvalue $\psi(s)$ must lie between $-\eee^{-\beta E_N}$ and 0~\footnote{In writing equation~\eqref{finiteN}, we have assumed that $N$ is still large. A more precise calculation that does not neglect terms of $O(1/N)$ shows that the exact eigenvalue equation for a finite number of configurations is $\frac{1}{N} \sum_{i = 1}^N \frac{\eee^{-E_i/T}}{\psi(\g)+c \eee^{-E_i/T}} = \eee^{\g}$ with $c = (N - 1 + \eee^{-s})/N$. As can be seen, this correction does not affect the conclusions for large $N$ as $c\to 1$.}. Because $E_N$ diverges for $N\to\infty$, then $\psi(s) \to 0$ in the limit. The implication is that $\g>0$ always produces an inactive dynamical phase, where the time--averaged activity $\langle {\bar{a}} \rangle_s=-\psi'(s)$ also goes to zero. Note that by extension of the above argument, the second largest eigenvalue of $\WW(\g)$ lies between $-\eee^{-\beta E_{N-1}}$ and $-\eee^{-\beta E_N}$. These two bounds also vanish for large $N$ and so there is no gap spectral gap in the thermodynamic limit. Thus to understand the dynamics in the inactive biased phase, we need to use the general formalism introduced in an earlier section, with the time-dependent weight factors $q(E,\tau)$.

To look at the time-dependent trap depth distribution $P_s(E,\tau)$ (equation~\eqref{P_C_tau_dependent}) for a given $\g$ and total trajectory length $t$ we need three ingredients: the forward probability $P_{\rm{f}}(E,\tau)$, the backward factor $q(E,\tau)=Q(E, t - \tau)$ and the normalizing partition function $Z(\g,t)$. If one Laplace transforms these with respect to the relevant time variables $\tau$, $t-\tau$ and $t$, respectively, explicit expressions can be found. As before we choose an initial state where the system finds itself in a randomly chosen trap, corresponding to $P_{\rm{f}}(E,0)=\rho(E)=\eee^{-E}$. The initial condition for the backward factor is, according to the discussion after equation~\eqref{equ:path_prob_multiplicative},  $q(E, t)  = Q(E,0) = 1$.

The forward (eq.~\eqref{pf_continuum}) and backward (eq.~\eqref{final_Qe}) equations in Laplace space are 
\begin{align}
  z \hat{P}_{\rm{f}}(E, z) - P_{\rm{f}}(E,0) &= -{\rm{e}}^{-E/T}  \hat{P}_{\rm{f}}(E,z) + {\rm{e}}^{-s} \rho(E) \int dE'\,{\rm{e}}^{-E'/T} \hat{P}_{\rm{f}}(E',z) \\
  z \hat{Q}(E, z) - Q(E,0) &= -{\rm{e}}^{-E/T}  \hat{Q}(E,z) + {\rm{e}}^{-s} {\rm{e}}^{-E/T} \int dE'\,\rho(E') \hat{Q}(E',z)
\end{align}
with corresponding solutions
\begin{align}
  \hat{P}_{\rm{f}}(E,z) &= \frac{{\rm{e}}^{-E}}{(z+{\rm{e}}^{-E/T})\tilde{f}(z)}
  \label{forwlowbou} \\
  \hat{Q}(E,z) &= 
\frac{1}{z+{\rm{e}}^{-E/T} } \left( 1 + \frac{\eee^{-s}f(z)}{\tilde{f}(z)} \frac{\eee^{-E/T}}{z}\right)
% + \frac{1}{z}\frac{\eee^{-s}f(z)}{\tilde{f}(z)}
  \label{backcomp}
\end{align}
with the abbreviation
\be 
\tilde{f}(z) = 1-{\rm{e}}^{-s} (1-f(z)).
\ee
In the unbiased case $s=0$ one has $\tilde{f}(z)=f(z)$ and the solution for $\hat{P}_{\rm{f}}$ agrees with the one for $\hat{P}$ in~\eqref{Phat_unbiased} as it must, while similarly $\hat{Q}(E,z)$ simplifies to $1/z$, i.e.\ $Q(E,\Delta t)=1$ as expected.

For the case with bias towards low activity of interest here, $\g>0$, the key difference is that $\tilde{f}(z)$ now has a finite limit $1-\eee^{-s}$ for $z\to 0$. Using this in~\eqref{forwlowbou} and transforming back to the time domain then gives directly for long times
\begin{align}
  P_{\rm{f}}(E,\tau) &\approx \frac{\eee^{-r(E)\tau}\eee^{-E}}{1-\eee^{-\g}}
\label{pf_inactive_deep}
\end{align}
with $r(E) = \eee^{-E/T}$ as before. For $\hat{Q}$, if we assume that the relevant traps are those with escape rate $r(E)$ of order $z$, the second term in brackets in~\eqref{backcomp} can be neglected for $z\to 0$ because $f(z)\to 0$, giving
\begin{align}
  q(E,\tau)  &\approx  \eee^{-r(E)(t-\tau)}
\label{q_inactive_deep}
\end{align}
The partition function is found from equation~\eqref{partif_c}, which in the thermodynamic limit reads
\begin{align}
  Z(s,t) = \int_0^\infty dE\,P_{\rm{f}}(E,t)
  \label{particon}
\end{align}
%with $P_{\rm{f}}$ the solution to the master equation defined by the biased operator $\WW(s)$ (more on this below) and initial condition $P(E, 0) = \rho(E)$.
The Laplace transform w.r.t.\ total trajectory length $t$ of this is, from~\eqref{forwlowbou}, 
\be
\hat{Z}(s,z) = \frac{f(z)}{z\tilde{f}(z)}
\label{Zhat} 
\ee 
Using~\eqref{f_small_z} one then sees that for long times
\begin{align}
  Z(\g,t) &\approx \frac{T\Gamma(T)}{1-\eee^{-\g}}t^{-T}
           \label{Z_Bouchaud}
\end{align}
(This result applies also for $T>1$, where the singular term from~\eqref{f_small_z} for $T<1$ is no longer the leading small $z$ contribution but still gives the dominant large $t$ behaviour.)
The complete distribution according to equation~(13) is then
\begin{align}
P_s(E, \tau) &\approx \frac{t^T}{T \Gamma(T)} \eee^{-E}\eee^{- \eee^{-E/T}  t} 
\label{P_E_tau_inactive_Bouchaud}
\end{align}
Something remarkable happens here: the forward factor $P_{\rm{f}}(E,\tau)$ and the backward factor $q(E,\tau)$ both depend on the time $\tau$ along the trajectory, but this dependence cancels from the overall trap depth distribution. The physical interpretation is that after a (large enough, $\tau\gg 1$) transient time the trap depth distribution no longer changes for the remainder of the trajectory. Intriguingly, this distribution is {\em independent} of the specific value of the bias $\g$, as long as it is positive: the $\g$ dependence of $\Pf(E,\tau)$ and $Z(\g,t)$ cancels from $P_s(E,\tau)$. 

Further intuition can be obtained by noting that $P_s(E,\tau)$ in (\ref{P_E_tau_inactive_Bouchaud}) is peaked at values of $E$ around $T\ln t$ because it can be written as a function of $\delta E = E-T\ln t$ alone. For values of $E$ around the peak one has $r(E)t  =  \eee^{-E/T}t \sim 1$, meaning that the system will only make a number of jumps of $O(1)$ within the entire duration of the trajectory. The implication is that low activity-trajectories are realized by the system rapidly descending into traps that are so deep that it remains essentially stuck for the rest of the time, the dynamics then being ``frozen''.

\begin{figure}
  \includegraphics[width=0.5\textwidth]{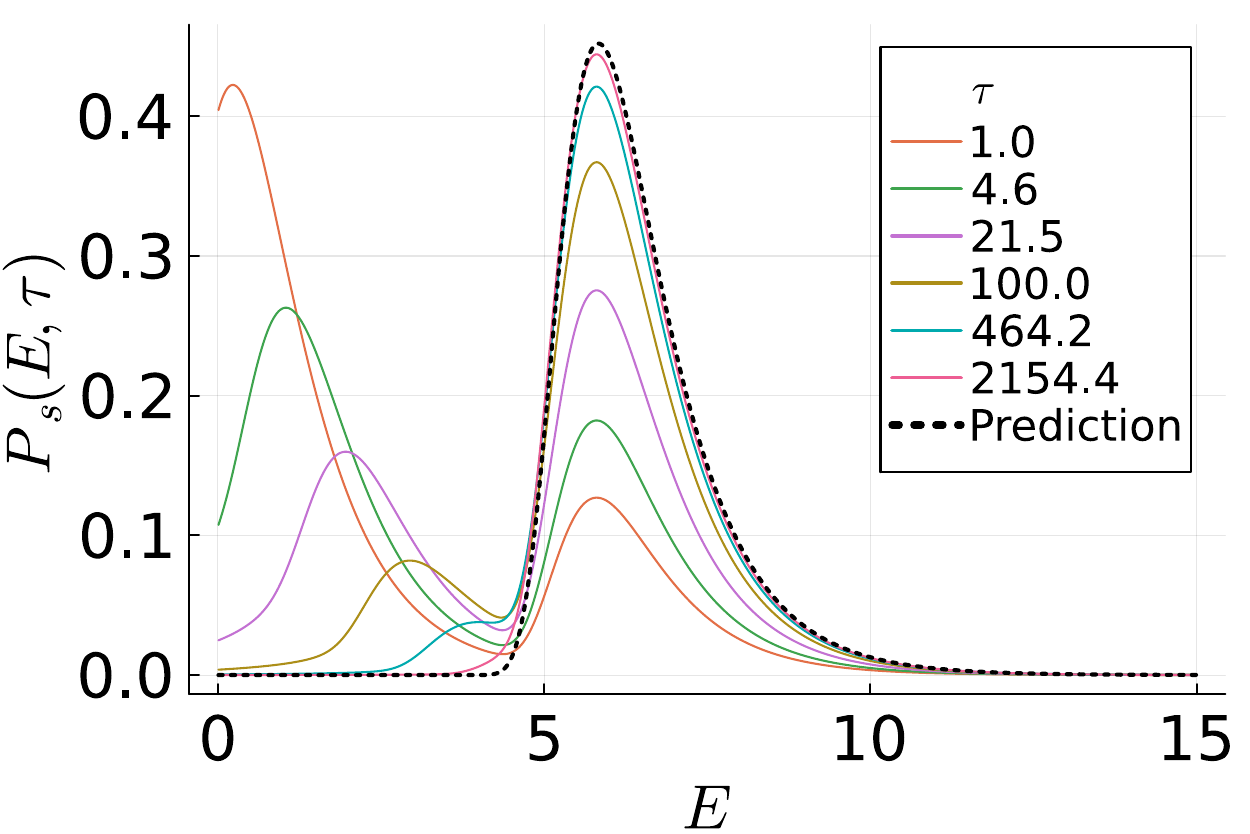}
  \caption{Trap depth distribution $P_s(E,\tau)$ in the Bouchaud trap model, at temperature $T=0.6$, bias parameter $\g=0.25$, total trajectory length $t=10^4$, evaluated at different running times $\tau$. Also shown is the prediction from equation~\eqref{P_E_tau_inactive_Bouchaud}.}
\label{fig:P_E_tau_dependence}
\end{figure}

The character of the low activity-dynamics can also be seen from numerical solutions for the time evolution of $P_s(E,\tau)$ as shown in Fig.~\ref{fig:P_E_tau_dependence}. Even at $\tau=1$ this distribution has a contribution proportional to the limiting form~\eqref{P_E_tau_inactive_Bouchaud}, because the bias prefers trajectories that already start in deep traps. As $\tau$ grows, the weight of the peak at $E$ of order unity is shifted into the large $\tau$-peak as trajectories starting from shallow traps rapidly descend.

We can substantiate the above picture by looking in more detail at the activity along biased trajectories. Considering first the mean long time--averaged activity, one finds from equations~(\ref{Z_Bouchaud}) and~(5) that 
\be
\langle \bar{a} \rangle_s = -\frac{1}{t} \partial_\g \ln Z(\g,t) = \frac{1}{t} \frac{1}{\eee^\g-1}
 \label{Bouchaud_total_act}
\ee
Correspondingly the total activity is
\begin{equation}
  \label{finite_activity}
  \langle \mathcal{A} \rangle_s = t \langle \bar{a}\rangle_s = \frac{1}{\eee^s - 1}
\end{equation}
consistent with our expectation that the system will only make a finite number of jumps along the entire trajectory. As an aside we note that this number diverges as $\g\to 0$ as it must, to connect with the extensive ($\sim t$) number of jumps that we found in the active phase ($\g<0$, see for instance Fig.~\ref{fig:free_energy_Bouchaud}).

To get more detailed information we next consider the mean time--dependent activity. This can be evaluated using equation~(12). Using the approximations~\eqref{q_inactive_deep} and~\eqref{P_E_tau_inactive_Bouchaud}, for which we assumed deep traps (with $e^{-E/T} \sim z$) to be dominant, would actually yield a vanishing result. This tells us that we need a more accurate treatment that includes the contribution from the shallow traps. To achieve this, we first rewrite~(12) as
\be 
\langle a_\tau\rangle_s = 
\int dE\, \left[r(E) + \frac{\partial_{\Delta t}Q(E,\Delta t)}{Q(E,\Delta t)} \right]
\frac{Q(E,\Delta t)P_{\rm{f}}(E,\tau)}{Z(s,t)}
=
\int dE\,
\frac{S(E,\Delta t)P_{\rm{f}}(E,\tau)}{Z(s,t)}
\label{act_tau_exact}
\ee
where
\be
S(E,\Delta t) =  r(E)Q(E,\Delta t) + \partial_{\Delta t}Q(E,\Delta t) 
\ee
Now from~\eqref{backcomp}, the Laplace transform of this function is
\be 
 \hat{S}(E,z) = \left(
1 + \frac{\eee^{-s}f(z)}{\tilde{f}(z)} \frac{\eee^{-E/T}}{z}\right) - Q(E,\Delta t=0) = \eee^{-s}\eee^{-E/T} \hat{\sigma}(z)
\ee
%where
%The first term is the Laplace transform of $\delta(\Delta t)$ (which arises formally from the fact that $Q(E,\Delta t)$ has a step from 0 to 1 at $\Delta t=0$ because of the initial condition $Q(E,\Delta t=0)=1$) and can be discarded as we are interested in $\Delta t>0$. The remaining term only depends multiplicatively on $\eee^{-E/T}$ so we can write $\hat{S}(E,z)=\eee^{-s}\eee^{-E/T}\hat{\sigma}(z)$ 
with
\be 
\hat{\sigma}(z) = 
\frac{f(z)}{z\tilde{f}(z)}
\label{sigmahat}
\ee
Carrying out the integration over $E$ in~\eqref{act_tau_exact} we then obtain for the mean time-dependent activity
\be 
\langle a_\tau\rangle_s = \frac{\sigma(\Delta t)\pi(\tau)}{Z(s,t)}
\ee
where 
\be 
\hat{\pi}(z) = \eee^{-s}\int dE\, \eee^{-E/T}\hat{P}_{\rm{f}}(E,z) = \eee^{-s}\,\frac{1-f(z)}{\tilde{f}(z)}
\label{pihat}
\ee
and $\sigma(\Delta t)$ and $\pi(\tau)$ are the inverse Laplace transforms of $\hat{\sigma}(z)$ and $\hat{\pi}(z)$, respectively. For $\sigma(\Delta t)$, we see by comparing~\eqref{Zhat} and~\eqref{sigmahat} that in fact $\sigma(\Delta t) = Z(s,\Delta t)$ and so from~\eqref{Z_Bouchaud} $\sigma(\Delta t)/Z(s,t) = (\Delta t/t)^{-T}$, which is unity if we are not near the end of the trajectory, i.e.\ as long as $\tau \ll t$ so that $\Delta t \approx t$. Moving on to $\pi(\tau)$, for long times $f(z)$ is small and we can expand
\be 
\hat{\pi}(z) = \eee^{-s}\frac{1-f(z)}{1-\eee^{-s}+\eee^{-s}f(z)} \approx \frac{1}{\eee^s-1} - f(z)\,\frac{\eee^s}{(\eee^s-1)^2}
\ee
The singular term from~\eqref{f_small_z} then determines the long-time behaviour (for $1\ll \tau\ll t$), giving for the mean time--dependent activity the final result
\be 
%the term $1/\tilde{f}(z)$ in~\eqref{pihat} is constant to leading order in $z$ so can be ignored. The remainder is $-f(z)/\tilde{f}(z)=-z\hat{Z}(s,z)$, giving in the time domain
%\be 
%\pi(\tau) = -\partial_\tau Z(s,\tau) = \frac{T^2\Gamma(T)}{1-\eee^{-s}}\tau^{-T-1}
%\ee
%Multiplying by the constant $\sigma(\Delta t)/Z(s,t) \approx \eee^{-s}$ gives then, finally, for the mean time--dependent activity
\langle a_\tau\rangle_s = \pi(\tau) = 
%\eee^{-s}\frac{T^2\Gamma(T)}{1-\eee^{-s}}\tau^{-T-1}= 
\frac{T^2\Gamma(T)\eee^{s}}{(\eee^s-1)^2}\tau^{-T-1}
\label{pos_bias_scaling}
\ee
The power law exponent here is $-(T+1)$ as stated in the main text. The prediction~\eqref{pos_bias_scaling} is tested against numerics in Fig.~\ref{activ_bou}, showing good agreement including the prefactor.

\subsection{Barrat--M\'ezard trap model}

In this section, we consider activity--biased dynamics in the Barrat--M\'ezard trap model. Because of the more complicated transition rates (eq.~(2)) obtaining analytical results is more difficult than for the Bouchaud case. Here, we focus on the regime that is qualitatively different from the Bouchaud trap model, namely temperatures $T < 1/2$ (cf.\ the phase diagrams in Fig.~1 in the main text).

We start as in the previous section with the calculation of the mean time--dependent activity in the unbiased case. Then, we set out the arguments that allow us to obtain the phase boundary of equation~(7). Finally, we calculate the mean time--dependent activity in the presence of a bias, at zero temperature, and discuss the validity of the resulting scaling across the whole robust aging phase (as shown e.g.\ in Fig.~2(b) in the main text).

\subsubsection{Zero bias}

The master equation~(3) with the appropriate transition rates (eq.~(2)) and corresponding escape rate $r(E)$ reads
\begin{align}
  \frac{\partial}{\partial \tau} P(E,\tau) &=  - 
%F(1,T,1+T,-{\rm{e}}^{E/T} ) 
r(E) P(E,\tau) + \rho(E) \int dE' \frac{P(E',\tau)}{1+{\rm{e}}^{-(E - E')/T} }
                                             \label{zerobias_bm}
\end{align}
with initial condition $P(E,0) = \rho(E) = {\rm{e}}^{-E}$. The escape rate is 
\be 
r(E) = \int dE'\,W(E\to E')\rho(E') = \int dE' \frac{\eee^{-E'}}{1+{\rm{e}}^{-(E' - E)/T} }
= \int dE' \frac{\eee^{-E'}\eee^{-E/T}}{\eee^{-E/T}+{\rm{e}}^{-E'/T} }  = f(\eee^{-E/T})
\ee
with $f(z)$ as defined in~\eqref{f_def}. In the long time limit, typical trap depths are large and we can focus on the leading behaviour~\eqref{f_small_z} of $f(z)$ for small arguments, giving
\begin{align}
r(E)  
%F(1,T,1+T,-{\rm{e}}^{E/T} )  
\approx \frac{\pi T}{\sin(\pi T)} {\rm{e}}^{-E}
  \label{ratebm2}
\end{align}
%and hence we may expand the hypergeometric function for large values of $E$. T
This is proportional to the fraction $\eee^{-E}$ of traps of depth greater than that of the initial trap, which is consistent with the slow dynamics being governed by entropic barriers. The inverse of $r(E)$ yields the corresponding entropic timescale%
%This yields a simple variable to which compare the timescales
~\cite{sollich2006trap} and it is then natural, 
% The leading term of the expansion is
following Ref.~\cite{bertin2003cross}, to introduce the scaling variable  $u^{-1}  = \tau {\rm{e}}^{-E}$. The behavior for large and small $u$ thus corresponds to deep and shallow traps, respectively.
% and conversely, the behavior for small $u$ corresponds to shallow traps. 
It turns out that in terms of $u$ an asymptotic solution of equation~\eqref{zerobias_bm} can be found (see details in~\cite{bertin2003cross, sollich2006trap}). When converted back to the physical variables this solution reads
\begin{align}
  P(E,\tau) &\sim \tau {\rm{e}}^{-E} \, ,  \quad  &  \eee^E /\tau  \gg 1 \\
  P(E,\tau) &\sim \tau^{1-1/T} \eee^{-E} {\rm{e}}^{E/T} \, ,  \quad &  \eee^E /\tau  \ll 1
\label{shallow_bm}
\end{align}
We comment briefly on the regime $1/2 < T < 1$ where activation effects dominate~\cite{bertin2003cross, cammarota2015spontaneous}. One then finds, in analogy with the situation in the Bouchaud model, that the mean time--dependent activity is dominated by the shallow traps, i.e.\ by the part of the distribution corresponding to equation~\eqref{shallow_bm}. With this we can compute the scaling of the mean time--dependent activity as 
\begin{align}
  \langle a_\tau \rangle_0 &= \int dE\, P(E,\tau) r(E) \sim \tau^{1-1/T}
                           \label{actiscalingbm}
\end{align}
This was the result used in Fig.~2(a). 

For the regime $T<1/2$ that we mostly focus on in this section one finds, on the other hand, that the mean time--dependent activity is governed by typical traps with $\eee^{E} \sim \tau$~\cite{bertin2003cross, sollich2006trap}. This results in $\langle a_\tau\rangle_0 \sim r(E) \sim 1/\tau$, with an exponent that no longer depends on temperature $T$.

\subsubsection{Dynamical phase boundary for $T<1/2$}

  Here we give the arguments to derive equation~(7), which describes the phase boundary of the continuous dynamical phase transition at $T < 1/2$ (cf.\  phase diagram in Fig.~1(b)). We tackle the problem by approaching the boundary from the active side, i.e.\ $s < s^*$. 

\begin{figure}
  \centering
  \includegraphics[width=0.5\textwidth]{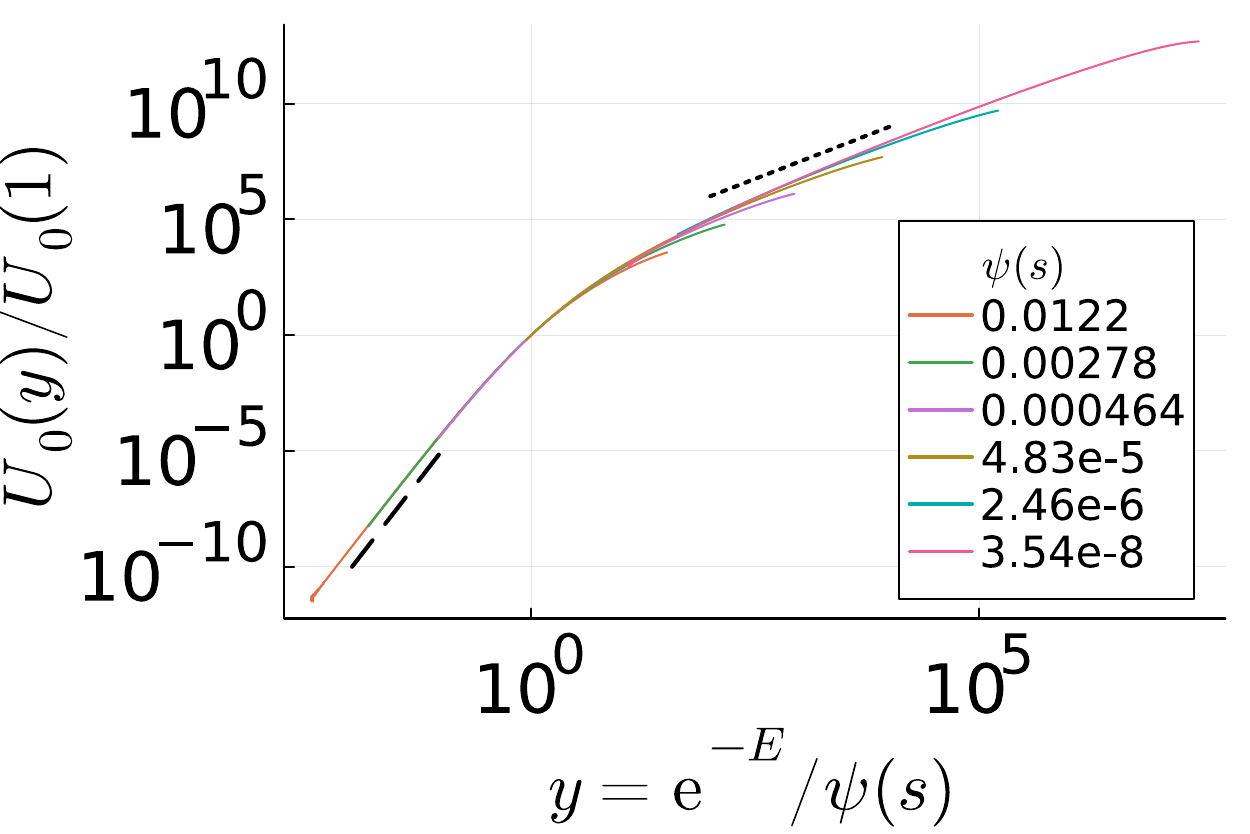}
  \caption{Normalized left (top) eigenvector components $U_0(E)$ as a function of the reduced variable $y  = x/\psi(s)=\eee^{-E}/\psi(s)$ obtained from the numerical solution of equation~\eqref{leftph} for different values of $s$ close to the phase boundary $s^*$. The resulting dynamical free energies are shown in the legend. Temperature: $T = 0.2$. Collapse of the curves into a single (master) curve indicates correctness of the assumption $U_0(x) \propto g(x/\psi(s))$. Dashed line: $g(y) \sim y^\beta$. Dotted line: $g(y) \sim y^{\beta/2 - 1}$.}
  \label{scalingu0}
\end{figure}

  Our starting point is the left eigenvector equation~\eqref{contin_lefte}, with the transition rates~(2). We write it as follows
  \begin{align}
    -r(E) U_0(E) + \eee^{-s} \int dE' \frac{ \rho(E') U_0(E')}{1 + {\rm{e}}^{-(E' - E)/T} } = \psi(s) U_0(E)
    \label{leftph}
  \end{align}
  with $r(E) \approx \frac{\pi T}{\sin(\pi T)} {\rm{e}}^{-E}$ (eq.~\eqref{ratebm2}) and $\rho(E) = \eee^{-E}$. With the change of variables $x = {\rm{e}}^{-E}$ equation~\eqref{leftph} becomes
  \begin{align}
    \eee^{-s} \int_0^1 dx' \frac{ U_0(x')}{1 + (x'/x)^\beta } =  U_0(x) (\psi(s) + c_T x)
    \label{leftx}
  \end{align}
  with $c_T = \frac{\pi T}{\sin(\pi T)}$ and $\beta = 1/T$. The form of the r.h.s.\ of~\eqref{leftx} suggests a solution with the scaling $U_0(x)\propto g(x/\psi(s))$ with $g$ a master curve to be determined; numerical results supporting this scaling form are shown in Fig.~\ref{scalingu0}. Inserting the assumed scaling and changing variables once more to $y=x/\psi(s)$ we get
  \begin{align}
    \eee^{-s} \int_0^{1/\psi(s)} dy' \frac{g(y') }{1+ (y'/y)^\beta} = g(y) (1+c_Ty)
    \label{inte}
  \end{align}
and at the phase boundary the upper integration boundary can be taken to infinity. Figure~\ref{scalingu0} suggests that the resulting $g(y)$ exhibits a crossover between two different power laws at $y=1$. The exponent we find numerically for small $y$ is consistent with the value $\beta$. Indeed, evaluating the two sides of~\eqref{inte} to leading order for small $y$ gives
  \begin{align}
    y^\beta \eee^{-s} \int_0^{\infty} dy' \frac{g(y')}{y'{}^\beta} = g(y)
\label{small_y}
  \end{align}
and hence $g(y)\sim y^\beta$. Of course the remaining integral on the l.h.s.\ has to be finite for this conclusion to hold: we find below that it is.

For large $y$ the exponent is nontrivial and we next establish the equations required to determine it. We rewrite eq.~\eqref{inte} for $y \gg 1$ as
  \begin{align}
     \eee^{-s} \int_0^ {\infty} dy' \frac{g(y') }{1+ (y'/y)^\beta} \approx g(y) c_Ty
  \end{align}
and insert the power law scaling $g(y) \sim y^\alpha$, with the exponent $\alpha$ to be determined, leading to
 % This leads us to the  following equation
  \begin{align}
    c_T \eee^s = \int_0^\infty dy' \frac{y'^{\alpha}}{1 + y'^\beta}
    \label{firsteq}
  \end{align}
This first condition relates $\alpha$ and $s$, but does not yet determine the position $s^*$ of the phase boundary. 

For the second equation we use a physical argument, based on an analogy with the unbiased case~\cite{cammarota2015spontaneous}. There, the dynamics changes at $T=1/2$ from being dominated by activation across effective energy barriers to slowing down by entropic barriers. This change is related to the fact that for $T>1/2$, the system is more likely to move towards shallower traps in each jump, and so eventually to the top of the landscape; while for $T<1/2$ it moves on average towards deeper traps. At $T=1/2$, the probabilities of moving to shallower and to deeper traps,
\be 
P_\uparrow(E) = \frac{1}{r(E)} \int_0^E dE' \,W(E \to E') \rho(E') \quad \mbox{and} \quad
P_\downarrow(E) = \frac{1}{r(E)} \int_E^\infty dE'\,W(E \to E') \rho(E')
\ee
are exactly equal (for deep traps, $E\gg 1$) as one can check by explicit calculation using $\rho(E')=\eee^{-E'}$.
%
%Starting from the active steady state and considering deep traps, i.e. \ $E \gg 1$, the probability of jumping towards the top of the landscape, %which is proportional to the rate of jumping to a shallower trap,
%$P_\uparrow(E) \propto \int_0^E dE' W(E \to E') {\rm{e}}^{-s} U_0(E')/U_0(E)$ is greater than the probability of going down 
Based on the picture that at the dynamical phase boundary for $T<1/2$ the driving is just strong enough to stop the system from aging, i.e.\ to allow it to return to shallow traps, we now posit that the analogous criterion holds there with the appropriate weighting by the left eigenvector $U_0(E')$, so that
\be 
P_\uparrow(E)  \propto \int_0^E dE'\,W(E \to E') \rho(E') U_0(E')
\quad \mbox{and} \quad 
P_\downarrow(E)  \propto \int_E^\infty dE'\,W(E \to E') \rho(E') U_0(E')
\ee 
should again be equal. 
%, with $U_0(E)$ the left top eigenvector of the biased master operator, which is essentially the backwards factor in the steady state~\footnote{See eqns.~\eqref{partisteady}--\eqref{equ:TTI_results} in supplemental material}. This makes aging unfeasible as the dominant behavior consists of jumps to shallower traps. That property must crossover at the phase boundary where we hypothesize that $P_\uparrow(E) = P_\downarrow(E)$. 
%To get an auxiliar relation, let us consider the escape from deep traps in the landscape of order ${\rm{e}}^{-E}/\psi(s) \sim 1$.  As discussed in the manuscript, the boundary should be characterized by the equality $ P_\uparrow(E) =  P_\downarrow(E)$ for the deepest traps. 
In terms of the scaling function $g(y)$ this equality reads, by analogy with the l.h.s.\ of~\eqref{inte}, 
\begin{align}
  \int_y^{\infty} dy' \frac{g(y')}{1 + (y'/y)^\beta} = \int_0^y dy'\frac{g(y')}{1+ (y'/y)^\beta}
\end{align}
or in terms of $w=y'/y$,
\begin{align}
   \int_1^{\infty} dw\,\frac{g(wy)}{1 + w^\beta} = \int_0^1 dw\,\frac{g(wy)}{1+w^\beta}
\end{align}
Finally we take the limit $y\to\infty$ because at the dynamical phase transition, where $\psi(s)\to 0$, all traps effectively have large $y=\eee^{-E}/\psi(s)$. With the assumed power law scaling $g(wy)\sim (wy)^\alpha$ for large $y$ our condition then becomes
\begin{align}
    \int_1^{\infty} dw \, \frac{w^\alpha}{1 + w^\beta} = \int_0^1 dw\,\frac{w^\alpha}{1+ w^\beta}
\end{align}
giving as the non--trivial solution for the exponent $\alpha$
\begin{align}
  \alpha = \frac{\beta}{2} -1
  \label{exprel}
\end{align}
This prediction for the exponent determining the large-$y$ behaviour of $g(y)$ is fully consistent with our scaling collapse of the numerical eigenproblem solutions in Fig.~\ref{scalingu0}. It also ensures that the integrand in~\eqref{small_y} scales as $y'{}^{-\beta/2-1}$ for large $y'$ and so the integral converges.
% in the temperature range $T<1/2$
%The validity of the scaling $y^\alpha$ with $\alpha$ given by equation~\eqref{exprel} for $y \gg 1$ is shown .
Finally, inserting the exponent value~\eqref{exprel} into equation~\eqref{firsteq} and solving for $s$ one finds expression~(7) in the manuscript for the location of the dynamical phase boundary.

\subsubsection{Zero temperature}

In the unbiased model, we saw above  that the decay of the time-dependent activity at $T < 1/2$ has a $T$-independent power law; in fact the physics of the slow dynamics here is driven by an entropic mechanism~\cite{bertin2003cross, sollich2006trap} that exists down to zero temperature. We therefore now focus on this limit and solve for the scaling of the mean time--averaged activity at $T = 0$, allowing for any value of the bias (including the undriven case $s=0$). We find that the activity continues to scale as $1/\tau$, consistent with our numerical observation that this scaling applies for any $T < 1/2$ in the robust aging phase.

At zero temperature, the transition rates (eq.~(2)) reduce to Heaviside functions ($\Theta(x) = 0$ for $x < 0$ and  $\Theta(x) = 1$ for $x \geq 0$) and the escape rate becomes the  (complement of the) cumulative distribution of the density of states, namely  $r(E) = \int_0^\infty dE' \Theta(E'-E) \rho(E') = \int_E^\infty dE' \rho(E')$. In what follows, we will consider a general distribution $\rho(E)$ because for zero temperature and zero bias it is known that the results are in fact independent of $\rho(E)$~\cite{barrat1995phase}.

We write for this scenario the forward (eq.~\eqref{pf_continuum}) and backward (eq.~\eqref{final_Qe})  equations as
\begin{align}
  \frac{\partial}{\partial \tau} P_{\rm{f}}(E, \tau) &= -r(E) P_{\rm{f}}(E, \tau) + \eee^{-s} \rho(E) \int_0^E dE'  P_{\rm{f}}(E', \tau)
  \label{partforw} \\
  \frac{\partial}{\partial \Delta t} Q(E, \Delta t) &= -r(E) Q(E, \Delta t) + \eee^{-s} \int_E^\infty dE'  \rho(E') Q(E', \Delta t)
                                                                  \label{partback}
\end{align}
with $\Delta t = t - \tau$ as before. It will be convenient to transform those equations from trap depths $E$ to the corresponding escape rates $r(E)$. We take into account that $P_{\rm{f}}(E, \tau) $ transforms as a probability distribution, i.e.\ $P_{\rm{f}}(E, \tau)  = P_{\rm{f}}(r, \tau)  \left| \frac{d r}{d E} \right|=  P_{\rm{f}}(r, \tau) \rho(E)$; whereas $Q(E, t - \tau) $ transforms as a scalar function. Hence the transformed equations are
\begin{align}
   \frac{\partial}{\partial \tau} P_{\rm{f}}(r, \tau) &= -r  P_{\rm{f}}(r, \tau) + \eee^{-s} \int_r^1 dr' P_{\rm{f}}(r', \tau)  \\
  \frac{\partial}{\partial \Delta t} Q(r, \Delta t) &= -r Q(r, \Delta t) + \eee^{-s} \int_0^r d r' \, Q(r', \Delta t) 
\end{align}
Now we Laplace transform the equations, using the initial conditions $P_{\rm{f}}(r, 0) = \frac{P_{\rm{f}}(E, 0)}{\rho(E)} = 1$ and $Q(r, 0) = 1$, to yield
\begin{align}
\hat{P}_{\rm{f}}(r, z) {(z+r)}  &=   1 + \eee^{-s}  \int_r^1 d r' \hat{P}_{\rm{f}}(r', z) \label{for1} \\
  \hat{Q}(r, z)(z + r)  &= 1 + \eee^{-s} \int_0^r d r'  \,  \hat{Q}(r', z)  \label{back1}
\end{align}
These equations have the explicit solutions
%For those equations, the reader can verify the following solutions 
\begin{align}
  \hat{P}_{\rm{f}}(r, z) &= \frac{(z+1)^{b}}{(z+r)^{b+1}}
  \label{pf0} \\
  \hat{Q}(r, z) &= \frac{(z+r)^{b-1}}{z^b}
                       \label{q0}
\end{align}
with $b = {\rm{e}}^{-s}$. The partition function in Laplace space follows as
\begin{align}
  \hat{Z}(s, z) = \int_0^1 d r \, \hat{P}_{\rm{f}}(r, z) = \frac{(z+1)^b}{b z^b} - \frac{1}{b}
  \label{z0}
\end{align}

\begin{figure}
    \centering
    \includegraphics[width=0.5\textwidth]{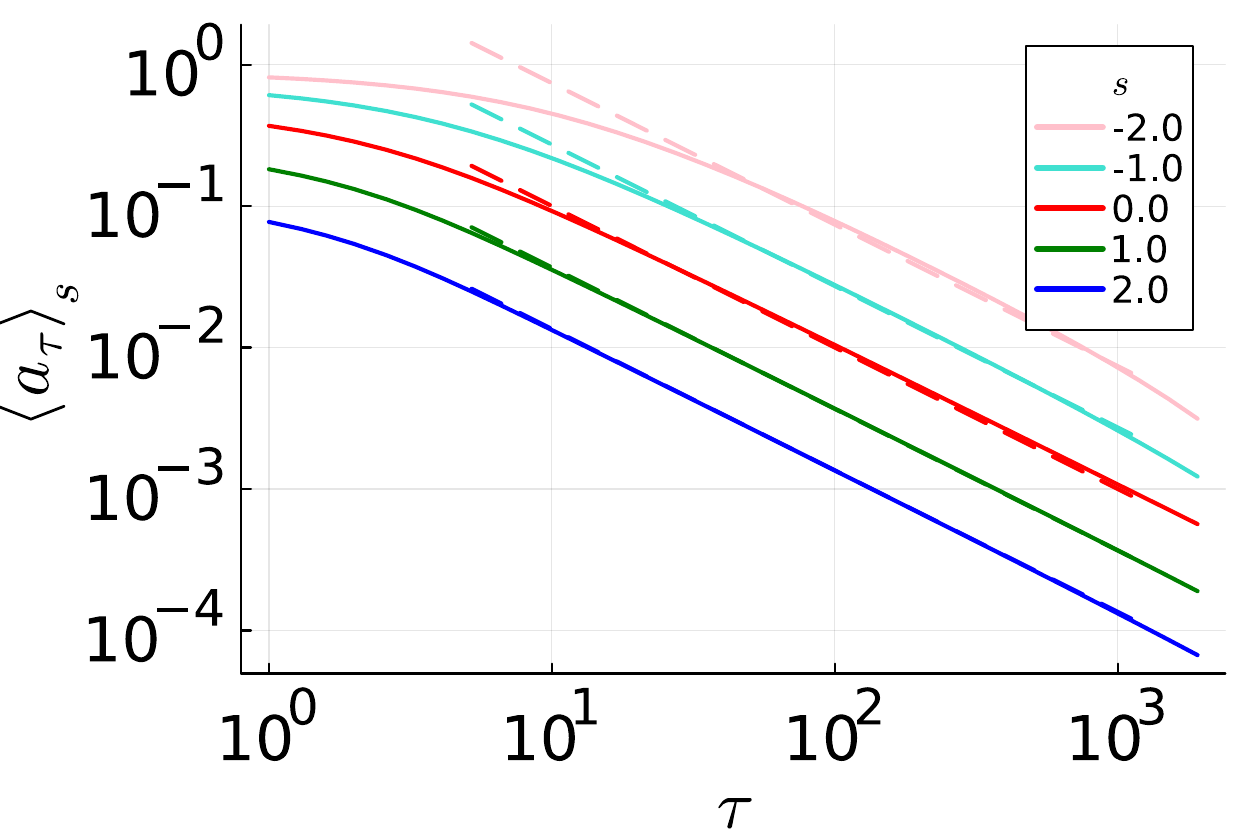}
    \caption{Mean time--dependent activity (eq.~(12)) for $T = 0$ obtained via the numerical integration of equations~\eqref{partforw} and~\eqref{partback} (solid lines) compared against the scaling prediction~\eqref{final0} (dashed lines), for different biases $s$ as shown. Total trajectory length $t = 10^{5}$.}
    \label{zeroT}
  \end{figure}

We now focus as usual on the long-time limit $z\ll 1$, where we expect the dynamics to concentrate around deep traps $r\ll 1$. In $\hat{P}_{\rm{f}}(r, z)$ the singularity with the largest real part is at $z=-r$. The numerator $(z+1)^b$ is unity there to leading order so we can approximate $P_{\rm{f}}(r, z)\approx (z+r)^{-b-1}$. For $\hat{Q}(z)$ the dominant singularity is at $z=0$ so that to leading order $\hat{Q}(r, z) = r^{b-1}z^{-b}$. Using the analogous argument for $\hat{Z}(s,z)$ and carrying out the inverse Laplace transforms we obtain
%For long times, we consider that 
%$z \ll r \ll 1$. The inverse transforms then become tractable and yield 
% This approximation is used to inverse Laplace transform relations~\eqref{pf0}--\eqref{z0}. 
%This 
\begin{align}
  P_{\rm{f}}(r, \tau) &\approx \frac{{\rm{e}}^{-r \tau} \tau^b }{\Gamma(1+b)} \\
  q(r, \tau) &\approx \frac{(t-\tau)^{b-1} r^{b-1}}{\Gamma(b)} \\
  Z(s, t) &\approx \frac{t^{b-1}}{\Gamma(1+b)}
  \label{Z_BM}
\end{align}
{and hence
\begin{align}
  P_s(r, \tau) \approx   \frac{{\rm{e}}^{-r \tau}   r^{b-1} \tau^b (t-\tau)^{b-1} }{\Gamma(b) t^{b-1}}
\end{align}
}
The auxiliary escape rate can now be calculated from eq.~(12) and becomes $r+(b-1)/(t-\tau)$. The correction term is negligibly small as long as we are far from the end of the trajectory,
% far from the right boundary, 
i.e.\ $\tau \ll t$.
% reveals a  contribution from the backwards factor of order $t^{-1}$ that turns negligible for the $\tau$--scaling. 
We therefore have
\begin{align}
  \langle a_\tau \rangle_s &\approx \int_0^1 d r \, P_s(r, \tau)  r \\
                           &\approx \frac{(t - \tau)^{b-1} b}{\tau t^{b-1}} \\
                           &\approx \frac{\eee^{-s}}{\tau}
\label{final0}
\end{align}
assuming again $\tau\ll t$.
In Figure~\ref{zeroT} we test this prediction against numerical solutions of the forward and backward equations, and see that both the asymptotic scaling and its prefactor are captured correctly.
%predict correctly the asymptotics.
The mean total activity is from~\eqref{Z_BM}
\be
\langle \mathcal{A}\rangle_s = - \partial_s \ln Z(s,t) = \eee^{-s}\ln t 
\label{BM_activity_zerotemp}
\ee
to leading order for large $t$. This is consistent with~\eqref{final0}; it also shows that the system does not freeze, i.e.\ that the mean number of jumps diverges with $t$. {This property characterizes the behavior of the whole robust aging phase that we find for $T < 1/2$. We remark that the limit $s \to \infty$ is somewhat singular as the prefactor in (\ref{BM_activity_zerotemp}) vanishes; indeed this limit produces strictly zero activity at any finite $t$. Nonetheless the activity diverges for $t\to\infty$ at any finite $s>0$, including for large $s$ when the (non-commuting) limits are taken in the physically relevant order $\lim_{s\to\infty}\lim_{t\to\infty}\langle \mathcal{A}\rangle_s$.
%, one needs to take first $t \to \infty$ to obtain results compatible with any finite positive $s$
} The fact that the system continues to age can also be seen from the explicit form of the time-dependent distribution across traps (again for $\tau\ll t$)
\be
P_s(r,\tau) \propto P_{\rm{f}}(r,\tau) q(r,t-\tau) \propto \eee^{-r\tau} (r\tau)^{b-1}
\ee
This shows simple aging scaling, i.e.\ a dependence only on $r(E)\tau$, so that the distribution shifts to deeper and deeper traps as $\tau$ increases.

In the main text, in Fig.~2(b), we show that the scaling from eq.~\eqref{final0} is consistent also with numerical results for finite temperatures in the robust aging phase. This supports our conjecture that the scaling $\langle a_\tau\rangle \sim \tau^{-1}$
%Comparing equations~\eqref{actiscalingbm} and~\eqref{final0}, we see that at $T = 1/2$ the scalings agree, and thus we think that (i) the scaling becomes
is independent of temperature throughout this phase, with both temperature and bias only affecting the prefactor as for the case $T = 0$.

\subsection{First order perturbation theory}

In this section, we explain how we obtain the mean time--averaged activity as shown in e.g.\ Fig.~1(d) {(see also Fig.~\ref{fig:free_energy_Bouchaud}(right)) }, without having to take numerical derivatives of the dynamical free energy $\psi(s)$. We work for simplicity with a finite number $N$ of traps $\C$ and exploit the fact that $\WW(s)$ can be symmetrized by a similarity transformation involving the diagonal matrix associated with the formal equilibrium (Boltzmann) distribution for $s = 0$, i.e.\ ${\mathbb{P}}_{\rm{eq}} \propto {\rm{diag}} ({\rm{e}}^{\beta E_\C})$ (see also the comments after~\eqref{contin_lefte}). This equilibrium solution always exists for finite $N$, though in the glassy regime the time to reach it diverges with $N$.

Explicitly, the similarity transformation reads
\begin{align}
  \tilde{\WW}(s) = {\mathbb{P}}_{\rm{eq}}^{-1/2}  \WW(s) {\mathbb{P}}_{\rm{eq}}^{1/2}
  \label{symW}
\end{align}
or in components
\begin{align}
  \tilde{W}_{\C \Cp}(s)  = {\rm{e}}^{-\beta E_\C/2} W_{\C \Cp}(s)  {\rm{e}}^{\beta E_\Cp/2} \, .
\end{align}
Denoting by $\tilde{V}_0$ the normalised eigenvector corresponding to the top eigenvalue of $\tilde{\WW}(s)$, we have
\begin{align}
\psi(s) =  \langle \tilde{V}_0 | \tilde{\WW}(s) |\tilde{V}_0 \rangle 
\end{align}
One can now deploy standard first--order perturbation theory (as used for instance in Ref.~\cite{de2016rare}; the key feature is that the $s$-dependence of $\tilde{V}_0$ does not contribute) to express the first derivative of the free energy as
\begin{align}
  \psi'(s) &= \langle \tilde{V}_0 | \tilde{\WW}'(s) |\tilde{V}_0 \rangle  \\
  &= \sum_{\C, \Cp} \tilde{V}_{0}(\C) \tilde{W}'_{\C \Cp}(s) \tilde{V}_{0}(\Cp) \\
   &= \sum_{\C, \Cp} \tilde{V}_{0}(\C)   {\rm{e}}^{-\beta E_\C/2} W'_{\C \Cp}(s)  {\rm{e}}^{\beta E_\Cp/2} \tilde{V}_{0}(\Cp) 
\end{align}
Now we use  that the eigenvectors of the original biased master operator can be written in terms of the symmetrized eigenvector as $\langle U_0 | = \langle \tilde{V}_0 | {\mathbb{P}}_{\rm{eq}} ^{-1/2}$ and $| V_0 \rangle =  {\mathbb{P}}_{\rm{eq}} ^{1/2} | \tilde{V}_0 \rangle $ (see discussion after~\eqref{contin_lefte} and  Ref.~\cite{margiotta2018spectral} for an illustration in the case of the unbiased Bouchad trap model). Additionally, we use the expression of the master operator as given by equation~\eqref{equ:WA}. This results in
\begin{align}
   -\psi'(s) = \sum_{\C, \Cp, \C \neq \Cp} U_{0}(\C) W(\Cp \to \C) {\rm{e}}^{-s} V_{0}(\Cp)
   \label{firsto}
\end{align}
and in the thermodynamic limit %, eq.~\eqref{firsto} reads
\begin{align}
  % \psi'(s) &= -\sum_{E, E', E \neq E'} \langle U(E) | \mathbb{W}(s) | V(E') \rangle \\  &\approx -\int dE \int dE' U_0(E) V_0(E') W(E' \to E) {\rm{e}}^{-s} \rho(E)
 -\psi'(s) &= \int dE \,dE' \,U_0(E) V_0(E') W(E' \to E) {\rm{e}}^{-s} \rho(E) 
\label{cont_first}
\end{align}
{One can easily verify that the r.h.s. of equation~\eqref{cont_first} corresponds to the mean time--averaged activity by using the equation for the top right eigenvector (eq.~\eqref{rightE}) to replace the integral over $E'$. Then equation~\eqref{cont_first} transforms to
\begin{align}
  -\psi'(s) &= \int dE \,U_0(E) V_0(E) (\psi(s) + r(E) ) 
\end{align}
and one can recognize this as equation~(14) of the manuscript.
}

For Barrat--M\'ezard transition rates, eq.~\eqref{cont_first} becomes
\begin{align}
  -\psi'(s) = \int dE\,dE'\,\frac{U_0(E) V_0(E') \rho(E)   {\rm{e}}^{-s}}{1+ {\rm{e}}^{-\beta(E- E')}} = \langle \bar{a} \rangle_s
  \label{longacti}
\end{align}
and this expression was used to generate the data for Fig.~1(d) in the main text.

%For the numerical calculations the most efficient way to obtain the top left and right eigenvectors is in fact from the solution of the eigenvector equation for the symmetrized operator $\tilde{\WW}(s)$ (eq.~\eqref{symW}). Its form in the thermodynamic limit can be found from e.g.~\eqref{contin_lefte} by noting that the equilibrium distribution across trap depths is $\propto \rho(E)\eee^{\beta E}$ so that the eigenvectors transform as $V_0(E)\propto \rho(E)^{1/2}\eee^{\beta E/2}\tilde{V}_0(E)$ and $U_0(E)\propto \rho(E)^{-1/2}\eee^{-\beta E/2}\tilde{V}_0(E)$. Inserting into~\eqref{contin_lefte} then gives for the symmetrized problem 
  %\begin{align}
  %  -r(E) \tilde{V}_0(E) + \eee^{-s} \int  dE'\,\frac{[\rho(E)\rho(E')]^{1/2}}{2 \cosh \left(\beta( E - E') /2 \right)}\,\tilde{V}_0(E') = \psi(s) \tilde{V}_0(E) 
 % \end{align}
%Then one can use the equilibrium distribution $\propto \rho(E)\eee^{\beta E}$ to transform the solutions accordingly, i.e.\ $V_0(E)=\rho(E)^{1/2}\eee^{\beta E/2}\tilde{V}_0(E)$ and $U_0(E)=\rho(E)^{-1/2}\eee^{-\beta E/2}\tilde{V}_0(E)$.

\subsection{Numerical details}

{Here we provide a brief summary of how we generated the main figures of the manuscript. The dynamical free energy in Fig.~1(c) for the active regime  is obtained as the top eigenvalue of a  biased master operator $\WW(\g)$ of size $N = 2^{13}$. An instance of the master operator with elements given by eq.~\eqref{equ:WA} can be generated after randomly sampling $N$ energies from to $\rho(E)$ and associate to each configuration one of these energies. To use first--order perturbation theory for the determination of the mean time--averaged activity we obtain the top right eigenvector using the Arnoldi iteration (as implemented in the Arpack package~\cite{arpack}) and then exploit detailed balance to obtain the top left--eigenvector (see discussion after eq.~\eqref{contin_lefte}). }

{The numerical calculation of the mean time--dependent activity  (see e.g. Fig.~2) proceeds in two steps. First, we find the biased probability distribution at a given time (eq.~(13), see e.g. Fig.~3 and Fig.~\ref{fig:P_E_tau_dependence}) and then we compute the mean auxiliary escape rate according to eq.~(12).} {To obtain $P_s(E, \tau) $  we start by discretizing the energy range $E \in [0, 15]$ with a grid of step size $\Delta E = 0.02$, we also fix the total time (or trajectory length) $t = 10^4$ and consider a discrete time step $\Delta \tau = 5 \times 10^{-2}$. Then we discretize the forward and backward master equations for $\WW(s)$ (eq.~\eqref{pf_continuum} and~(11), respectively), replacing the time derivatives by finite differences. Finally, we use the initial conditions $\Pf(E, 0) = {\rm{e}}^{-E}$ and $q(E, t) = 1$ and then combine the numerical solutions according to equation~(13). To evaluate the mean time--dependent activity (eq.~(12)) it then remains to estimate the derivative of the logarithm of the backwards factor. This is done on--the--fly during the numerical integration, using again finite differences. }

%\end{document}

%\printbibliography
%\bibliography{cavi}

%\putbib
%\end{bibunit}

\end{document}